\documentclass[prd,preprint,showpacs,preprintnumbers,amsmath,amssymb,eqsecnum]{revtex4-1}

\usepackage{dcolumn}
\usepackage{amsfonts,mathrsfs}
\usepackage{bm}
\usepackage{subfigure}
\usepackage{url}
\usepackage[dvipdfmx]{hyperref,graphicx,color}
\begin{document}

\preprint{RUP-17-27}

\title{Complete conformal classification 
of the Friedmann-Lema\^{i}tre-Robertson-Walker solutions 
with a linear equation of state}

\author{Tomohiro Harada$^{1}$}
\email{harada@rikkyo.ac.jp}
\author{B.~J.~Carr$^{2,3}$}
\author{Takahisa Igata$^{1}$}
\affiliation{$^{1}$Department of Physics, Rikkyo University, Toshima,
Tokyo 171-8501, Japan}
\affiliation{$^{2}$Department of Physics and Astronomy, Queen Mary University
of London, Mile End Road, London E1 4NS, United Kingdom\\
$^{3}$Research Center for the Early Universe, 
Graduate School of Science, 
University of Tokyo, Tokyo 113-0033, Japan}
\date{\today}
\begin{abstract}
We completely classify Friedmann-Lema\^{i}tre-Robertson-Walker solutions
 with spatial curvature $K=0,\pm 1$ and equation of state $p=w\rho$,
 according to their conformal structure, singularities and trapping
 horizons. We do not assume any energy conditions and allow $\rho < 0$,
 thereby going beyond the usual well-known solutions. For each spatial
 curvature, there is an initial spacelike big-bang singularity for
 $w>-1/3$ and $\rho>0$, while no big-bang singularity for $w<-1$ and
 $\rho>0$. For $K=0$ or $-1$, $-1<w<-1/3$ and $\rho>0$, there is an
 initial null big-bang singularity. For each spatial curvature, there is
 a final spacelike future big-rip singularity for $w<-1$ and $\rho>0$,
 with null geodesics being future complete for $-5/3\le w<-1$ but
 incomplete for $w<-5/3$. For $w=-1/3$, the expansion speed is
 constant. For $-1<w<-1/3$ and $K=1$, the universe contracts from
 infinity, then bounces and expands back to infinity. For $K=0$, the
 past boundary consists of timelike infinity and a regular null
 hypersurface for $-5/3<w<-1$, while it consists of past timelike and
 past null infinities for $w\le -5/3$. For $w<-1$ and $K=1$, the
 spacetime contracts from an initial spacelike past big-rip singularity,
 then bounces and blows up at a final spacelike future big-rip
 singularity. For $w<-1$ and $K=-1$, the past boundary consists of
 a regular null hypersurface. The trapping
 horizons are timelike, null and spacelike for $w\in (-1,1/3)$, $w\in
 \{1/3, -1\}$ and $w\in (-\infty,-1)\cup (1/3,\infty)$, respectively. A
 negative energy density ($\rho <0$) is possible only for $K=-1$. In
 this case, for $w>-1/3$, the universe contracts from infinity, then
 bounces and expands to infinity; for $-1<w<-1/3$, it starts from a
 big-bang singularity and contracts to a big-crunch singularity; for
 $w<-1$, it expands from a regular null hypersurface and contracts to
 another regular null hypersurface.
\end{abstract}


\maketitle

\tableofcontents

\newpage

\section{Introduction}
\label{sec:introduction}
In the recent development of cosmology in general relativity 
and modified theories of gravity, 
the origin and the final fate of the universe
have been discussed intensively. 
In the context of general relativity, 
matter fields that violate some of the energy conditions, such as 
quintessence~\cite{Zlatev:1998tr} and 
phantom matter \cite{Caldwell:1999ew}, 
are proposed to explain the observed 
accelerated expansion of the universe. 
On the other hand, 
in a large class of modified theories of
gravity, the effective matter content can be defined by equating the effective stress-energy tensor 
with the Einstein tensor, in which case this 
 tensor may violate 
the energy conditions even if the 
matter stress-energy tensor does not. 
In both contexts, several curious scenarios for the origin and 
fate of the universe have recently been discussed, such as the 
genesis~\cite{Creminelli:2010ba} and 
big-rip models~\cite{Caldwell:2003vq}. 
The spacetime structure
in such 
scenarios is not always
well known -- but see Refs.~\cite{Dabrowski:2003jm,Dabrowski:2004hx,Chiba:2005er,FernandezJambrina:2006hj,Nemiroff:2014gea} -- 
and the purpose of this paper is to elucidate 
them.
 
If we assume a spatially homogeneous and isotropic spacetime, it is
uniquely given by the Friedmann-Lema\^{i}tre-Robertson-Walker (FLRW)
solution. The Einstein equations then require the effective 
matter content to be a perfect fluid. For a perfect
 fluid with 
equation of state $p=w\rho$,
where
$p$ and $\rho$ are the pressure and energy density, respectively, 
one may
impose the following conditions~\cite{Wald:1984rg}:
the null energy condition, $(1+w)\rho\ge 0$;
the weak energy condition, $\rho\ge 0$ and 
$(1+w)\rho\ge 0$;
the dominant energy condition, 
$\rho\ge |w\rho|$;
and the strong energy condition, $(1+w)\rho\ge 0$ and $(1+3w)\rho\ge 0$.
Observationally, the parameter $w$ is constrained to
$w=-1.00^{+0.04}_{-0.05}$ as the equation of state for dark 
energy responsible for the late-time acceleration of 
the Universe
under the assumption of $w$ being constant
~\cite{Abbott:2017wau}. 
It is important to know the conformal structure of the FLRW
spacetimes when
some of these energy conditions are violated.
The conformal structure of the FLRW solutions with $w\ge 0$ are already 
well known~(e.g. Refs.~\cite{Sato:1996,Senovilla:1998,Harada:2004pe}).
In the current paper, assuming the equation of state $p=w\rho$
but none of the energy conditions, we completely 
classify the FLRW solutions of the Einstein equations according to their 
conformal structure, singularities and trapping horizons.

The classification of the FLRW spacetimes is also motivated by 
models for the formation of black holes in 
asymptotically flat spacetimes and expanding universes.
Oppenheimer and Snyder~\cite{Oppenheimer:1939ue} used the 
FLRW spacetime with dust matched with a Schwarzschild exterior
to construct a model for
gravitational collapse to a black
hole.
Carr~\cite{Carr:1975qj} and Harada et al.~\cite{Harada:2013epa} used a 
positive-curvature FLRW 
interior and a flat FLRW 
exterior to construct models of primordial black hole formation.
Recently, Mena and Oliveira~\cite{Mena:2017kpo} 
used the FLRW spacetime with a generalised
Vaidya exterior to construct models for radiative gravitational 
collapse to an asymptotically anti-de Sitter black hole.
The conformal structure of solutions representing black holes in an expanding 
universe has been discussed in the context of 
the separate universe 
condition~\cite{Harada:2004pe,Harada:2005sc,Kopp:2010sh,Carr:2014pga}
and 
the  multiverse scenario~\cite{Deng:2016vzb,Garriga:2015fdk}.
In fact, this paper provides the background for a follow-up paper, 
in which we examine the conformal structure of solutions which represent black holes, wormholes and baby universes in a cosmological background.

This paper is organised as follows. We introduce basic concepts 
and present the conformal structures of maximally symmetric spacetimes in
Sec.~\ref{sec:preliminaries}. We classify the flat, positive-curvature and
negative-curvature FLRW solutions in Secs.~\ref{sec:flat},
~\ref{sec:PC} and~\ref{sec:NC},
respectively. We draw some general conclusions in Sec.~\ref{sec:conclusions}. 
Throughout this paper we use units in which $c=G=1$ and 
the abstract index notation~\cite{Wald:1984rg}.

\section{Preliminaries}
\label{sec:preliminaries}
\subsection{Spherically symmetric spacetimes, trapped
spheres  
and trapping horizons}
The line element in a spherically symmetric spacetime can be written in
the form
\begin{equation}
 ds^{2}=-e^{2\nu(t,r)}dt^{2}+e^{2\lambda(t,r)}dr^{2}+R^{2}(t,r)d\Omega^{2},
\end{equation}
where $d\Omega^{2}:=d\theta^{2}+\sin^{2}\theta d\phi^{2}$
is the line element on the unit 2-sphere.
The Misner-Sharp mass $m$ is defined as
\begin{equation}
 m:=\frac{R}{2}\left(1-\nabla_{a}R\nabla^{a}R\right)
\label{eq:MS}
\end{equation}
and
directly relates to the null expansions. 
If we define 
two independent null vectors which are orthogonal to a 
2-sphere with constant $t$ and $r$ by
\begin{equation}
 k^{a}_{\pm}=\frac{1}{\sqrt{2}}\left[e^{-\nu}\left(\frac{\partial
					      }{\partial t}\right)^{a}
\pm e^{-\lambda}\left(\frac{\partial}{\partial r}\right)^{a}\right],
\end{equation}
the associated null expansions are 
\begin{equation}
 \theta_{\pm}=\sqrt{2}\left(e^{-\nu}\frac{\partial}{\partial
		       t}\pm
		       e^{-\lambda}\frac{\partial}{\partial
		       r}\right)\ln R \, .
\end{equation}
In terms of $\theta_{+}$ and $\theta_{-}$, 
the Misner-Sharp mass can then be  rewritten in the 
 form~\cite{Nakao:1995ks}
\begin{equation}
 m=\frac{R}{2}\left(1+\frac{1}{2}\theta_{+}\theta_{-}R^{2}\right).
\end{equation}
A 2-sphere with constant $t$ and $r$ is called future trapped, past
trapped, future marginally trapped, past marginally trapped, bifurcating
marginally trapped and untrapped
if the signs of 
$(\max (\theta_{+}, \theta_{-}),\min (\theta_{+},\theta_{-}))$ 
are $(-,-)$, $(+,+)$,
$(0,-)$, $(+,0)$, $(0,0)$ and $(+,-)$, respectively. 
In this article, we simply call the closure of a hypersurface foliated by 
future (past) marginally trapped spheres a future (past)
trapping horizon~\cite{Hayward:1994bu}.
We call spacetime regions with 
null expansion signs
$(-,-)$, $(+,+)$ and $(+,-)$ future
trapped, past trapped and untrapped,
respectively.  
By definition, $R=2m$, $R<2m$ and $R>2m$ hold along the trapping
horizon, in the trapped region and in the untrapped region, respectively.  

\subsection{FLRW spacetimes, curvature tensor
and particle and event horizons}
For the FLRW spacetimes,
the metric is given by 
\begin{equation}
 ds^{2}=-dt^{2}+a^{2}(t)\left[dr^{2}+\Sigma_{K}^{2}(r)d\Omega^{2}\right],
\label{eq:FLRW_metric}
\end{equation}
where 
\begin{eqnarray}
 \Sigma_{K}(r)=\left\{\begin{array}{cc}
  r & (K=0)\\
  \sin r & (K=1) \\
  \sinh r & (K=-1)
		  \end{array}
\right. .
\label{eq:areal_radius}
\end{eqnarray}
For $K=0$, $1$ and $-1$, the spatial hypersurfaces are
flat, positive curvature and negative curvature, respectively.
In terms of the conformal time $\eta$, defined by 
\begin{equation}
 d\eta=\frac{dt}{a} \, ,
\end{equation}
the metric 
becomes conformal to the
static spacetime with constant spatial curvature:
\begin{equation}
 ds^{2}=a^{2}(\eta)\left[
-d\eta^{2}+dr^{2}+\Sigma_{K}^{2}(r)d\Omega^{2}\right].
\label{eq:FLRW_metric'}
\end{equation}

The components of the Ricci tensor are
\begin{eqnarray}
&& R_{00}=-3\left[\left(\frac{\dot{a}}{a}\right)^{\cdot}+\left(\frac{\dot{a}}{a}\right)^{2}\right],
  R_{0j}=0 \, , 
 R_{ij}=\left[\frac{2K}{a^{2}}+\left(\frac{\dot{a}}{a}\right)^{\cdot}+3\left(\frac{\dot{a}}{a}\right)^{2}\right]
 g_{ij} \, ,
\end{eqnarray}
where a dot denotes differentiation with respect to $t$ and 
$i$ and $j$ run over 1, 2, 3.
The Weyl tensor vanishes identically.
The Misner-Sharp mass for the FLRW spacetime is
\begin{equation}
 m=\frac{(a\Sigma_{K})^{3}}{2}\left(H^{2}+\frac{K}{a^{2}}\right),
\label{eq:MS_FLRW}
\end{equation}
where $H:=\dot{a}/a$. 
Since the null expansion pair is given by  
\begin{equation}
 \theta_{\pm}=\frac{\sqrt{2}}{a}\left(\dot{a}\pm \frac{\Sigma_{K}'}{\Sigma_{K}}\right) \, ,
\end{equation}
implying $\theta_+ + \theta_- = 2 \sqrt{2} \dot{a}/a$, a trapped sphere
with $\dot{a}>0$ ($<0$) is past (future) trapped.
This is also the case for marginally trapped spheres, trapped regions 
and trapping horizons. 

Particle and event horizons, unlike trapping horizons, are genuinely global
features of the spacetime. The future and past
event horizons of the observer with the world line $\gamma$ are defined 
as the boundaries of the causal past and future of $\gamma$,
respectively.
In cosmology, we usually adopt the inextendible
timelike geodesic of the isotropic observer as $\gamma$ 
and it can be shifted to $r=0$ by symmetry.
One conventionally  refers to future event horizons as cosmological event horizons but
particle horizons are conceptually different.
For an isotropic observer at 
an event $P$, whose position can be taken to be $r=0$, the particle horizon is the boundary between 
the world lines of comoving particles that can be seen by the 
observer at or before 
$P$ and 
those cannot be seen~\cite{Wald:1984rg}. 
From Eq.~(\ref{eq:FLRW_metric}), 
the comoving coordinate of the particle horizon 
for the event $P$ at time
$t=t_{P}$ is
\begin{equation}
 r_{\rm PH}(t_{P})=\lim_{\tau_{0}}\int^{t_{P}}_{\tau_{0}}\frac{dt}{a(t)} \, ,
\end{equation}
where the limit of $\tau_{0}$
is taken to be as small as possible 
within the spacetime. This 
means that the trajectory 
$(t,r_{PH}(t))$ gives an outgoing null hypersurface. Since this 
can be identified with the boundary of the causal future 
of $\gamma$ or the past event horizon of $\gamma$, 
the existence of particle horizons 
corresponds to the 
existence of past event horizons in the FLRW spacetimes. 
Note that this identification does not apply for general 
inhomogeneous spacetimes.

To determine the conformal structure of the spacetime, the
affine length of causal geodesics is important. By symmetry, the world line with constant $r$, $\theta$ and $\phi$
is a timelike geodesic, whose affine parameter 
is $t$ up to an affine transformation. 
The radial null geodesic is given by $\eta=\pm r+\mathrm{const}$.
By symmetry, any null geodesic with nonvanishing angular momentum 
can be shifted to a radial null geodesic 
by the spatial translation of coordinates. 
The affine parameter $\lambda$ of this null
geodesic can be given by
\begin{equation}
 \lambda=\int a^{2}d\eta=\int a(t)dt
\end{equation}   
up to an affine transformation.

\subsection{Dynamics and singularities in the FLRW spacetimes}
The Einstein equations require that the matter field be a
perfect fluid with
\begin{equation}
 T^{ab}=\rho u^{a}u^{b}+p(u^{a}u^{b}+g^{ab}) \, ,
\end{equation}
where $u^{a}$ is the 4-velocity, satisfying $u^{a}u_{a}=-1$,
and $\rho$ and $p$ are
the energy density and pressure, respectively.
The Einstein equations can be recast as the Friedmann equation,
\begin{equation}
 H^{2}=\frac{8\pi}{3}\rho-\frac{K}{a^{2}} \, ,
\label{eq:Friedmann}
\end{equation}
and the energy conservation equation,
\begin{equation}
 \dot{\rho}+3(\rho+p)H=0 \, .
\label{eq:energy_equation}
\end{equation}
Using Eqs.~(\ref{eq:MS_FLRW}) and (\ref{eq:Friedmann}), 
the Misner-Sharp mass 
is just
the energy density multiplied by a ``3-volume'': 
\begin{equation}
m= \frac{4\pi}{3}\rho (a\Sigma_{K})^{3} \, .
\end{equation}
Through the Einstein equations, the curvature invariants $R$ and
$R^{ab}R_{ab}$ can be written in terms of $\rho$ and $p$ as 
\begin{equation}
 R=8\pi(\rho-3p),\quad R^{ab}R_{ab}=64\pi^{2}(\rho^{2}+3p^{2}) \, .
\end{equation}

For equation of state $p=w\rho$, the energy conservation equation can be
integrated to give
\begin{equation}
 \rho= \rho_{0}\left(\frac{a_{0}}{a}\right)^{3(1+w)},
\label{eq:integration_conservation}
\end{equation}
where $\rho=\rho_{0}$ at $a=a_{0}$.
Thus the Friedmann equation can be  rewritten as
\begin{equation}
 H^{2}=\frac{8\pi}{3}\rho_{0}\left(\frac{a_{0}}{a}\right)^{3(1+w)}-\frac{K}{a^{2}} \, .
\label{eq:Friedmann1}
\end{equation}
For $w\ne -1/3$, this
can be transformed to
\begin{equation}
\dot{a}^{2} =\left(\frac{a_{c}}{a}\right)^{1+3w}-K,
\label{eq:Friedmann2}
\end{equation}
where 
\begin{equation}
 a_{c}^{1+3w}:=\frac{8\pi}{3}\rho_{0}a_{0}^{3(1+w)}.
\end{equation}
Through the coordinate transformation 
\begin{eqnarray}
 \tilde{a}=a^{1+3w},~~d\tilde{t}=(1+3w)\tilde{a}^{3w/(1+3w)}dt \, ,
\label{eq:tilde_variables}
\end{eqnarray}
this takes  the more familiar form 
\begin{equation}
 \left(\frac{d\tilde{a}}{d\tilde{t}}\right)^{2}=\frac{\tilde{a}_{c}}{\tilde{a}}-K,
\label{eq:Friedmann_dust}
\end{equation}
where 
\begin{equation}
 \tilde{a}_{c}:=\frac{8\pi}{3}\rho_{0}a_{0}^{3(1+w)}
\label{eq:tilde_a_c} 
\end{equation}
for $-\infty<w<\infty$. In terms of $\tilde{a}_{c}$, Eq.~(\ref{eq:MS_FLRW}) can be
rewritten as
\begin{equation}
 \frac{2m}{R}=\Sigma_{K}^{2}\frac{\tilde{a}_{c}}{a^{1+3w}} \, .
\end{equation}
For $w=-1/3$, Eq.~(\ref{eq:Friedmann1}) 
becomes
\begin{equation}
\dot{a}^{2} =\tilde{a}_{c}-K 
\label{eq:Friedmann3}
\end{equation}
and $2m/R$ is constant in time. 

There are
two types of 
singularities in the 
FLRW spacetimes considered here.
In the first type, as $t$ approaches some finite value from above (below), 
we have
$a\to 0$, $|\rho|\to \infty$ and $|H|\to \infty$, corresponding to
a big-bang (big-crunch) singularity.
In the second type, as $t$ approaches some finite
value from above (below), we have 
$a\to \infty$, $|\rho|\to \infty$ and $|H|\to \infty$, corresponding to
a future (past) big-rip singularity.
Both types of singularities occur  at a finite affine 
length along timelike geodesics.

\subsection{Maximally symmetric spacetimes}

For the vacuum  ($\rho =0$) case, 
Eq.~(\ref{eq:Friedmann}) has
no solution for $K=1$, while it gives $a=\mathrm{const}$ for $K=0$
and becomes
the standard Minkowski metric if one rescales $r$. 
As discussed  in Sec.~\ref{sec:NC},  
the empty FLRW solution for $K=-1$ is also a part of Minkowski
spacetime.
For $w=-1$, it can be seen from Eq.~(\ref{eq:energy_equation}) that $\rho$ and $p$
are constant.  
So the FLRW solution is 
de Sitter spacetime for $\Lambda>0$ and 
anti-de Sitter spacetime for
$\Lambda<0$, where $\Lambda = 8\pi \rho$ is the cosmological constant. These spacetimes are maximally symmetric.
Their conformal diagrams 
are 
well known but, following Ref.~\cite{Sato:1996}, 
we summarise them here for completeness.

\begin{figure}[htbp]
\begin{center}
\begin{tabular}{ccc}
\subfigure[Minkowski\label{fg:Minkowski}]{\includegraphics[height=0.3\textheight]{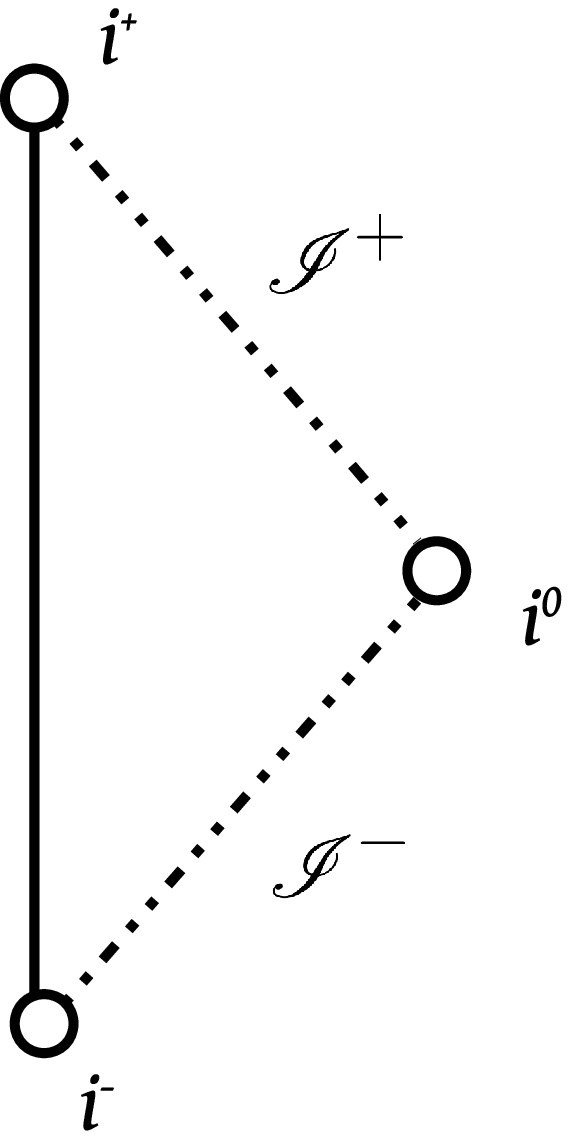}}&
\subfigure[dS\label{fg:deSitter}]{\includegraphics[height=0.25\textheight]{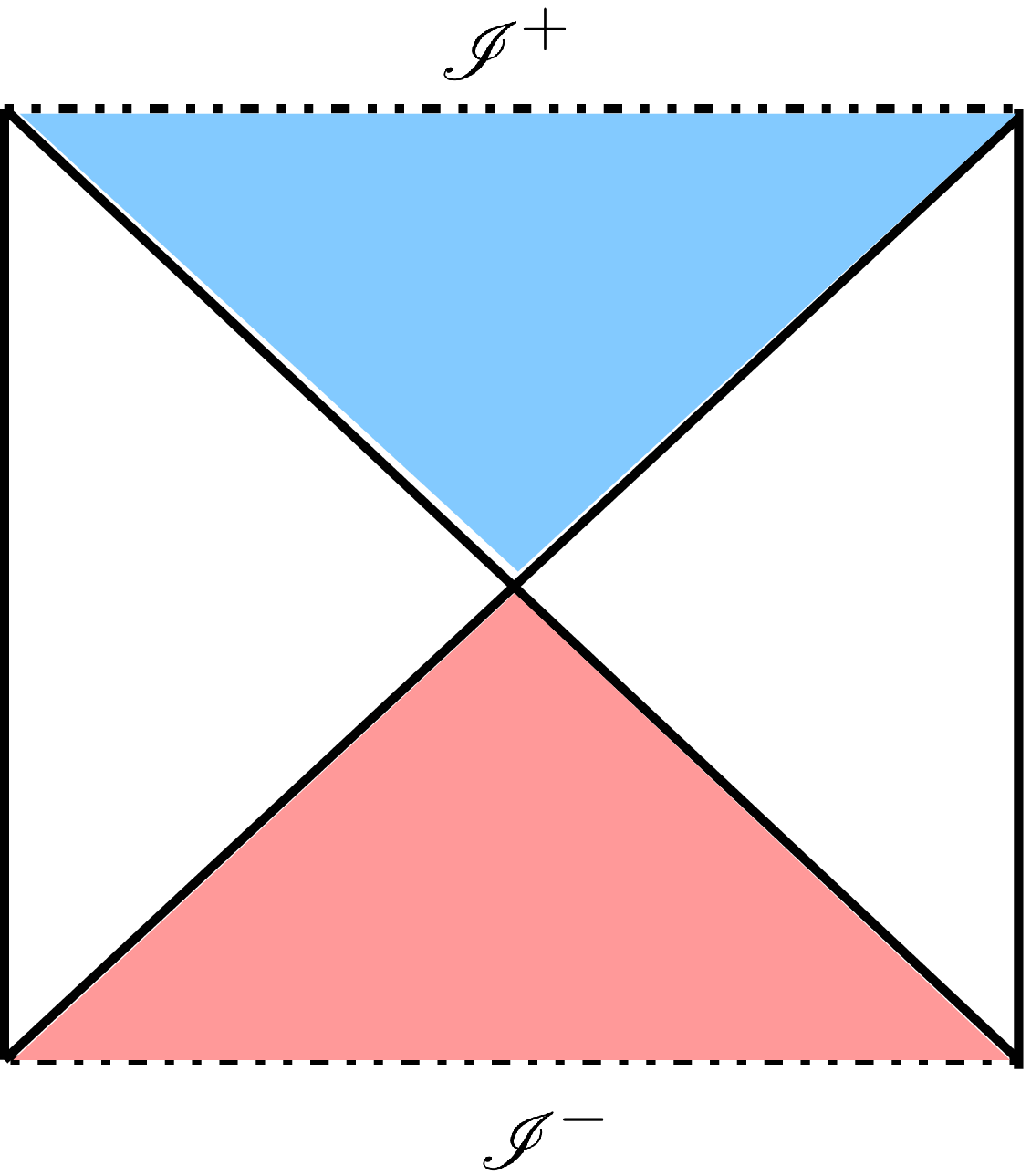}}&
\quad\quad \subfigure[AdS\label{fg:antideSitter}]
{\includegraphics[height=0.3\textheight]{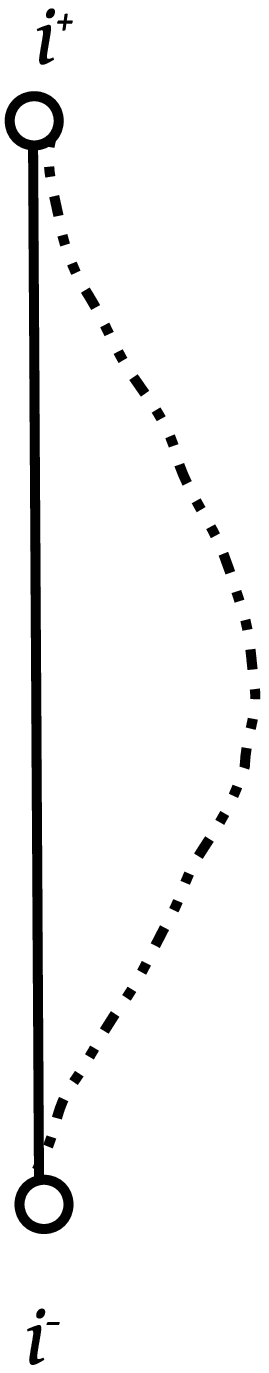}}
\end{tabular}
\caption{The conformal diagrams for the maximally symmetric spacetimes:
(a) Minkowski, (b) de Sitter and (c) anti-de Sitter.
The 
double-dotted 
lines
 denote 
infinities,
while the red and blue regions denote the future and past
 trapped regions, respectively.
The solid lines on the boundary of these regions
denote trapping horizons. 
}
\end{center} 
\end{figure}

\begin{figure}[htbp]
\begin{center}
\begin{tabular}{cccc}
\subfigure[Minkowski ($K=-1$)\label{fg:openMin}]{\includegraphics[height=0.24\textheight]{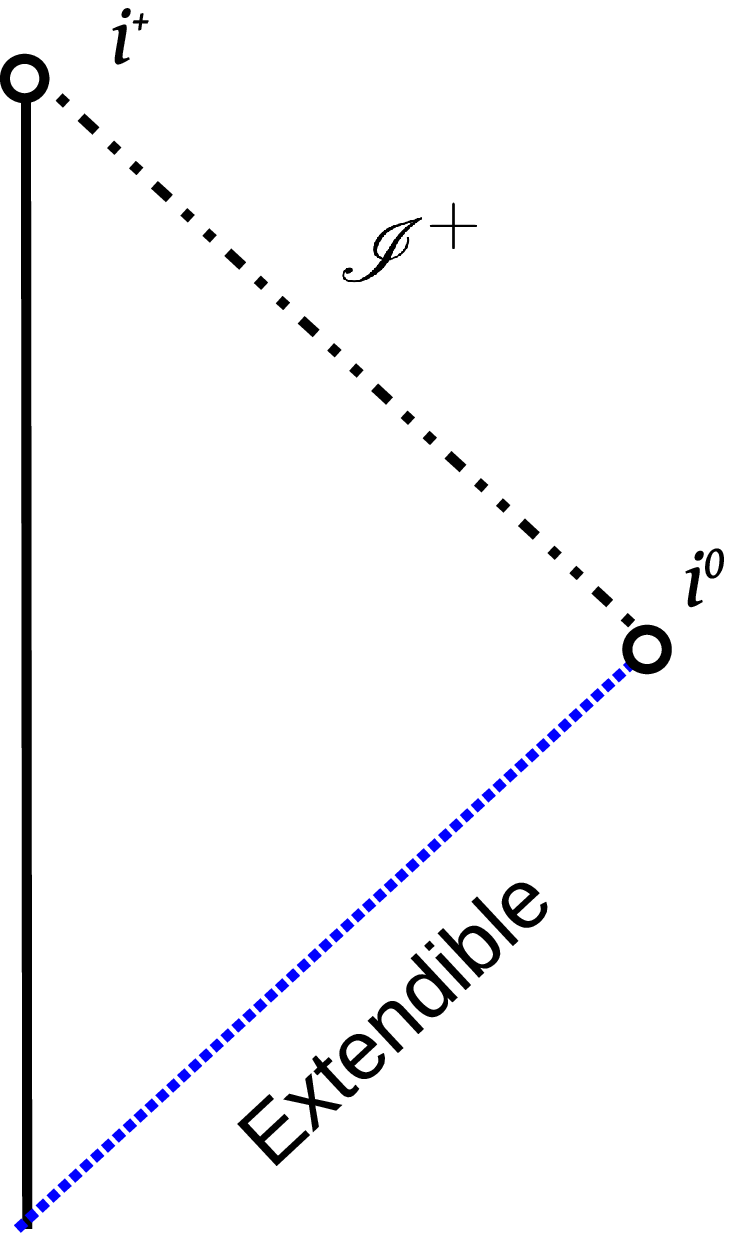}}&
\subfigure[dS ($K=0$)\label{fg:flatdS}]{\includegraphics[height=0.18\textheight]{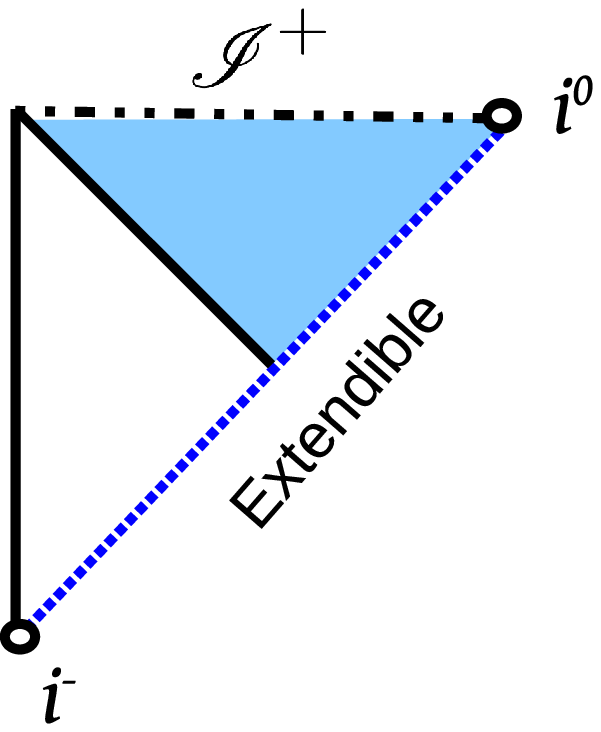}}&
\subfigure[dS ($K=-1$)\label{fg:opendS}]{\includegraphics[height=0.18\textheight]{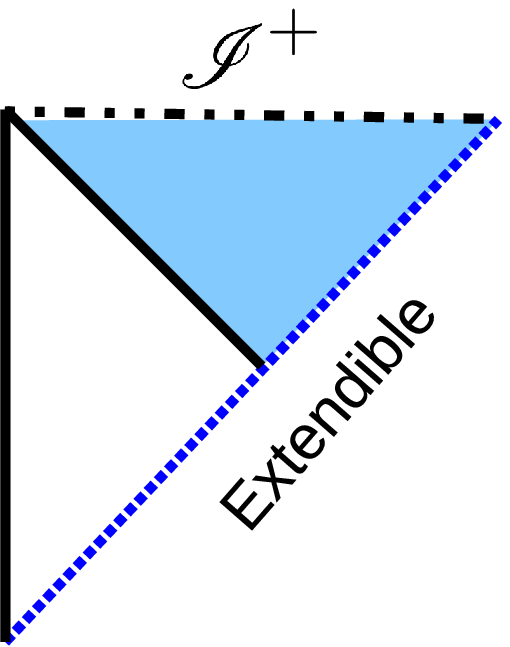}}&
\quad\quad \subfigure[AdS ($K=-1$)\label{fg:openAdS}]
{\includegraphics[height=0.24\textheight]{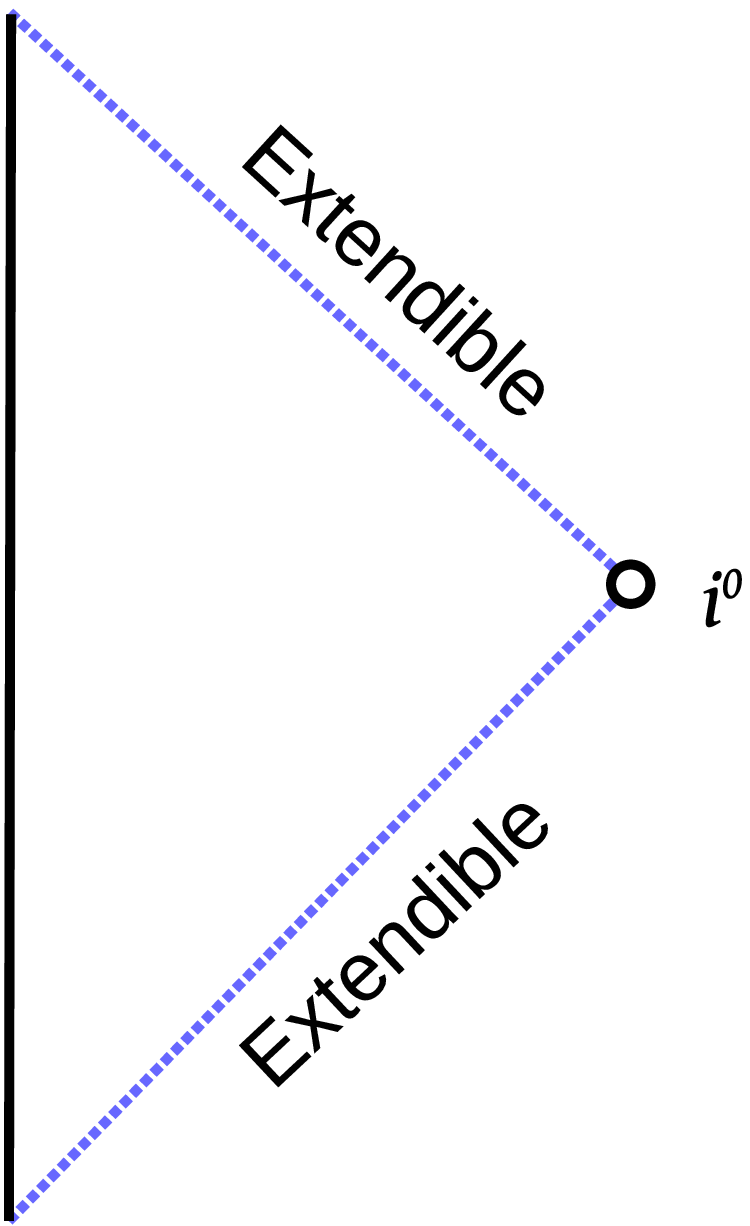}}
\end{tabular}
\caption{The conformal diagrams for 
(a) Minkowski in the open chart, (b) 
de Sitter in the flat chart, (c) de Sitter in the 
open chart and (d) anti-de Sitter in the open chart.
In all these cases, the chart covers only a 
part of the maximally extended spacetime.
}
\end{center} 
\end{figure}

\begin{itemize}
 \item Minkowski spacetime\\
The 
metric is
\begin{equation}
 ds^{2}=-dt^{2}+dr^{2}+r^{2}d\Omega^{2},
\end{equation}
where $-\infty<t<\infty$ and $r\ge 0$. 
This is the flat FLRW solution with $a=1$.
Using $\eta$ and $\chi$, where  
\begin{equation}
 2t=\tan \left( \frac{\eta+\chi}{2} \right) +\tan \left(\frac{\eta-\chi}{2} \right),\quad 
 2r=\tan \left(\frac{\eta+\chi}{2} \right) -\tan \left( \frac{\eta-\chi}{2} \right) \, ,
\end{equation}
with $|\eta+\chi|<\pi$, $|\eta-\chi|<\pi$ and $\chi\ge 0$, 
the metric can be transformed to 
\begin{equation}
ds^{2}=\frac{1}{4}\sec^{2} \left( \frac{\eta+\chi}{2} \right) \sec^{2} \left(\frac{\eta-\chi}{2} \right)
(-d\eta^{2}+d\chi^{2}+\sin^{2}\chi d\Omega^{2}) \, .
\end{equation}
Thus the Minkowski metric is conformal to 
Einstein's static
model,
\begin{equation}
ds^{2}_{\rm E} =-d\eta^{2}+d\chi^{2}+\sin^{2}\chi \, d\Omega^{2}.
\end{equation}
The (sectional)
conformal diagram 
for Minkowski spacetime is 
shown in 
Fig.~\ref{fg:Minkowski}. We see that 
$t=\pm \infty$ with $0\le r<\infty$ are mapped to $(\chi,\eta)=(0,\pm\pi)$,
corresponding to future and past timelike infinities,
while 
$\pm t=r=\infty$ are mapped to $\eta=\mp \chi\pm \pi$, 
corresponding to future and past null infinities.
The boundary $|t|<\infty$ and $r=\infty$ is mapped to 
$(\chi,\eta)=(\pi,0)$, corresponding to spatial infinity.
Since $2m/R=0$, there is no marginally trapped 
or trapped sphere.
There is no particle horizon or cosmological event horizon.
There is also 
an FLRW solution with $K=-1$ (i.e., in the open chart) given by 
\begin{equation}
 ds^{2}=-dt^{2}+t^{2}(dr^{2}+\sinh^{2}r d\Omega^{2}),
\end{equation}
in which 
the coordinate system only covers a part of the Minkowski spacetime.
This is sometimes called the Milne universe and 
its conformal diagram is shown in Fig.~\ref{fg:openMin}. 
This will be discussed in Sec.~\ref{sec:NC}.

\item de Sitter spacetime\\ 
The metric 
 in the global chart is 
\begin{equation}
 ds^{2}=\ell^{2}[-d\tau^{2}+\cosh^{2}\tau (dr^{2}+\sin^{2}r \, d\Omega^{2})] \, ,
\end{equation}
where $-\infty<\tau<\infty$, $0\le r\le \pi$  
and $\ell$ is a positive constant. This form of the metric 
corresponds to the FLRW spacetime with $K=1$.
The constant $\ell$ is related to $\Lambda$ by 
$
 \ell=\sqrt{3/\Lambda} \, .
$
The conformal time $\eta$ is given by 
$
 \eta=\arctan(\sinh \tau) \, ,
$
where the integration constant is chosen so that $-\pi/2<\eta<\pi/2$.
Hereafter
we will choose the integration constant without loss of generality so that 
the metric becomes
\begin{equation}
 ds^{2}=\ell^{2}\sec^{2}\eta \, ds^{2}_{{\rm E}} \, .
\end{equation}
The Misner-Sharp mass is given by 
\begin{equation}
 \frac{2m}{R}=\cosh^{2}\tau \sin^{2} r=\frac{\sin^{2}r}{\cos^{2}\eta} \, , 
\end{equation}
so the trapping horizon is 
$
 r={\pi}/{2}\pm \eta \, .
$
For $-\pi/2<\eta<0$, the region $\pi/2+\eta<r<\pi/2-\eta$ is future trapped;
for $0<\eta<\pi/2$, the region $\pi/2-\eta<r<\pi/2+\eta$ is past trapped.
The conformal diagram is shown in Fig.~\ref{fg:deSitter}.
There are both particle and cosmological event horizons.
There are also FLRW solutions with
      $K=0$ (i.e., in the flat chart) and $K=-1$ given by 
\begin{equation}
 ds^{2}=\ell^{2}[-d\tau^{2}+e^{2\tau}(dr^{2}+r^{2}d\Omega^{2})]
\end{equation}
and 
\begin{equation}
 ds^{2}=\ell^{2}[-d\tau^{2}+\sinh^{2}\tau (dr^{2}+\sinh^{2}r d\Omega^{2})],
\end{equation}
respectively, in both of which the coordinate system 
covers only a part of the spacetime. The conformal diagrams
are shown in Figs.~\ref{fg:flatdS} and \ref{fg:opendS}.

\item Anti-de Sitter spacetime\\
The metric in the universal covering space of anti-de Sitter spacetime
in the static chart is 
\begin{equation}
 ds^{2}=-\left(1+\frac{r^{2}}{\ell'^{2}}\right)dt^{2}+
\left(1+\frac{r^{2}}{\ell'^{2}}\right)^{-1}dr^{2}+r^{2}d\Omega^{2},
\end{equation}
where 
$
 \ell'=\sqrt{-{3}/{\Lambda}}
$
and 
$-\infty <t<\infty$ and $r\ge 0$.
In terms of $\eta$ and $\chi$, 
defined by  
$\eta=t/\ell'$ and $\tan\chi =r/\ell'$,
we have
\begin{equation}
 ds^{2}=\ell'^{2}\sec^{2}\chi \, ds_{{\rm E}}^{2} \, ,
\end{equation}
where $0\le \chi < \pi/2$. In terms of $\tilde{\eta}$ and
$\tilde{\chi}$,
defined by 
\begin{equation}
 2\eta=\tan \left( \frac{\tilde{\eta}+\tilde{\chi}}{2} \right) +\tan \left( \frac{\tilde{\eta}-\tilde{\chi}}{2} \right), ~~
 2\chi=\tan \left(\frac{\tilde{\eta}+\tilde{\chi}}{2} \right) -\tan \left(\frac{\tilde{\eta}-\tilde{\chi}}{2} \right),
\end{equation}
we find 
\begin{equation}
 ds^{2}
=\frac{\ell'^{2}}{4}\sec^{2}\chi
\sec^{2} \left( \frac{\tilde{\eta}-\tilde{\chi}}{2} \right)
\sec^{2} \left( \frac{\tilde{\eta}+\tilde{\chi}}{2}\right)
\left[-d\tilde{\eta}^{2}+d\tilde{\chi}^{2}+\sin^{2}\tilde{\chi}\frac{\sin^{2}\chi}{\chi^{2}}d\Omega^{2}\right].
\end{equation}
We note that the domain 
$-\infty<\eta<\infty$, $0\le \chi < \pi/2$ is transformed to
\begin{equation}
-\pi<\tilde{\eta}<\pi, \quad 0\le \tilde{\chi}< \pi, \quad  \tan \left( \frac{\tilde{\eta}+\tilde{\chi}}{2} \right)-
 \tan \left( \frac{\tilde{\eta}-\tilde{\chi}}{2} \right) < \pi .
\end{equation}
From
 Eq.~(\ref{eq:MS}) the Misner-Sharp mass is 
\begin{equation}
 m=-\frac{r^{3}}{2\ell'^{2}} \, .
\end{equation} 
Therefore
the whole region is untrapped, with no trapping horizon or trapped region.
The conformal diagram is 
shown in Fig.~\ref{fg:antideSitter}.
There is no particle horizon or cosmological event horizon.
There is also an FLRW solution with
      $K=-1$ given by 
\begin{equation}
 ds^{2}=\ell'^{2}[-d\tau^{2}+\cos^{2}\tau (dr^{2}+\sinh^{2}r d\Omega^{2})],
\end{equation}
in which the coordinate system covers only a part of the 
spacetime. The conformal diagram is shown in Fig.~\ref{fg:openAdS}.
\end{itemize}

\section{Flat FLRW solutions}
\label{sec:flat}
\begin{figure}[htbp]
\begin{center}
{\includegraphics[width=0.9\textwidth]{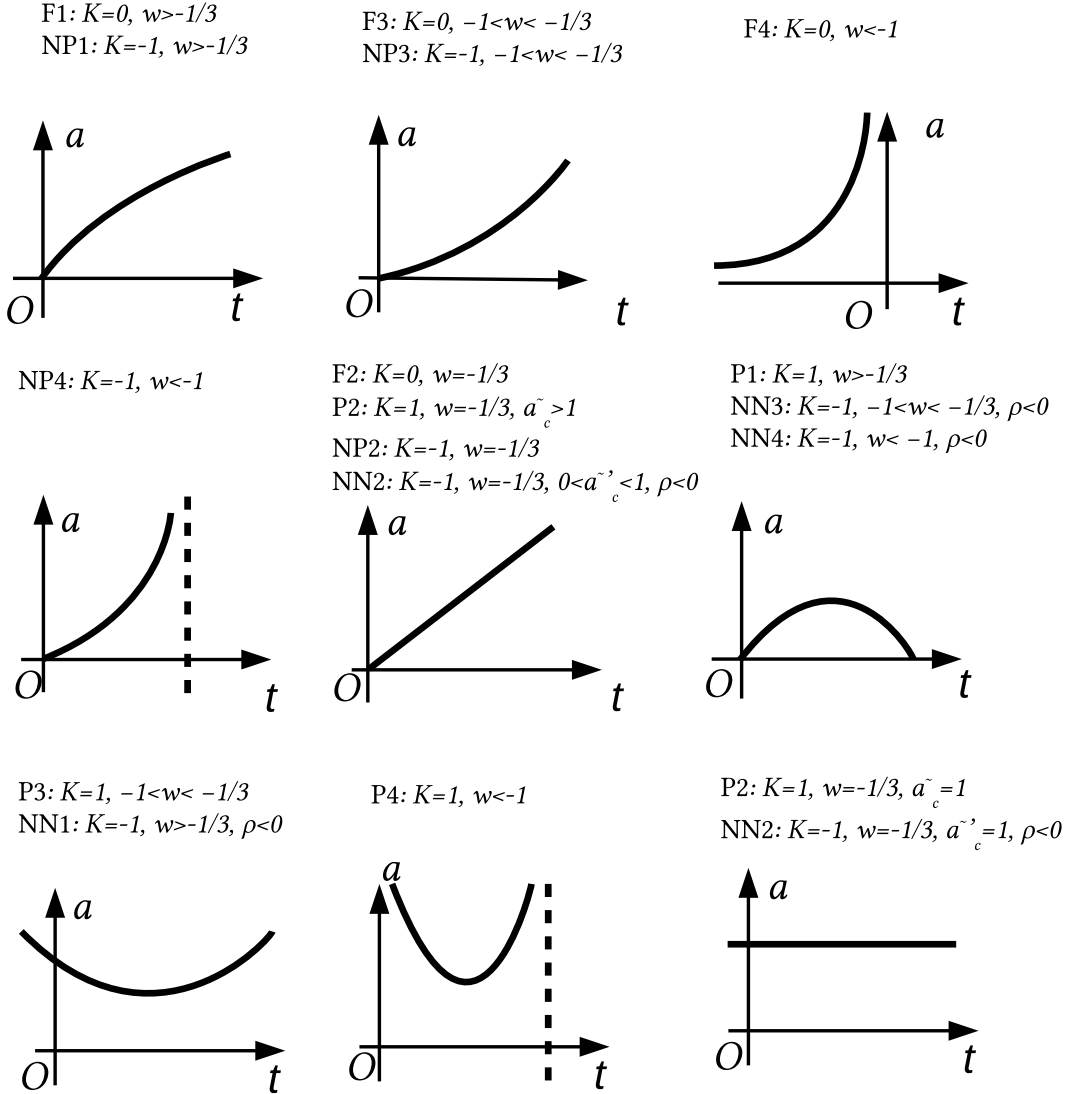}}
\caption{\label{scalefactor} Evolution of the scale factor $a(t)$ for
 different values of $K$, $w$ and the parameter describing the solutions.} 
\end{center} 
\end{figure}

As can be seen from Eq.~(\ref{eq:Friedmann2}), 
$a$ is a monotonic function of $t$ for $K=0$. 
We choose the expanding branch, 
the collapsing branch being
the time reverse of this.
Equation~(\ref{eq:Friedmann1}) can be integrated to give 
\begin{equation}
 a=C t^{\frac{2}{3(1+w)}}
\end{equation}
for $w>-1$ and $0<t<\infty$ or
\begin{equation}
 a=C (-t)^{\frac{2}{3(1+w)}}
\end{equation}
for $w<-1$ and $-\infty<t<0$, the constant being
\begin{equation}
 C=[6\pi (1+w)^{2}\rho_{0}]^{\frac{1}{3(1+w)}}a_{0} \, .
\end{equation}
The form of $a(t)$ is shown in Fig.~\ref{scalefactor} and we note that
the gradient is infinite (zero) at $t=0$ for $w<-1$ or
$-1/3< w$ ($-1< w < -1/3$).
The conformal time is 
\begin{equation}
 \eta=\frac{1}{C}\frac{3(1+w)}{1+3w}t^{\frac{1+3w}{3(1+w)}}
\end{equation}
for $w >-1$ (but excluding $w=-1/3$) and
\begin{equation}
 \eta=-\frac{1}{C}\frac{3(1+w)}{1+3w}(-t)^{\frac{1+3w}{3(1+w)}}
\end{equation}
for $w < -1$. Since the metric 
has the conformally flat form in terms of 
$\eta$, the conformal diagram for  this spacetime 
is just part of the 
Minkowski spacetime shown in Fig.~\ref{fg:Minkowski}
for some appropriate  
domain of $\eta$ and $r$. 
For $w\ne -1/3$, we have
\begin{equation}
 \frac{2m}{R}=\left(\frac{2}{1+3w}\right)^{2}\frac{r^{2}}{\eta^{2}}\, ,
\end{equation}
so there is a trapping horizon along 
\begin{equation}
 \eta=\pm \frac{2}{1+3w}r \, .
\end{equation}
The region $r>|(1+3w)\eta|/2$ is past trapped.
Note that there are just past trapped regions in this case.

\begin{figure}[htbp]
\begin{center}
\subfigure[\label{fg:flatwgm13} F1: $w> 1/3$, $w=1/3$,
 $-1/3<w<1/3$]{\includegraphics[height=0.2\textheight]{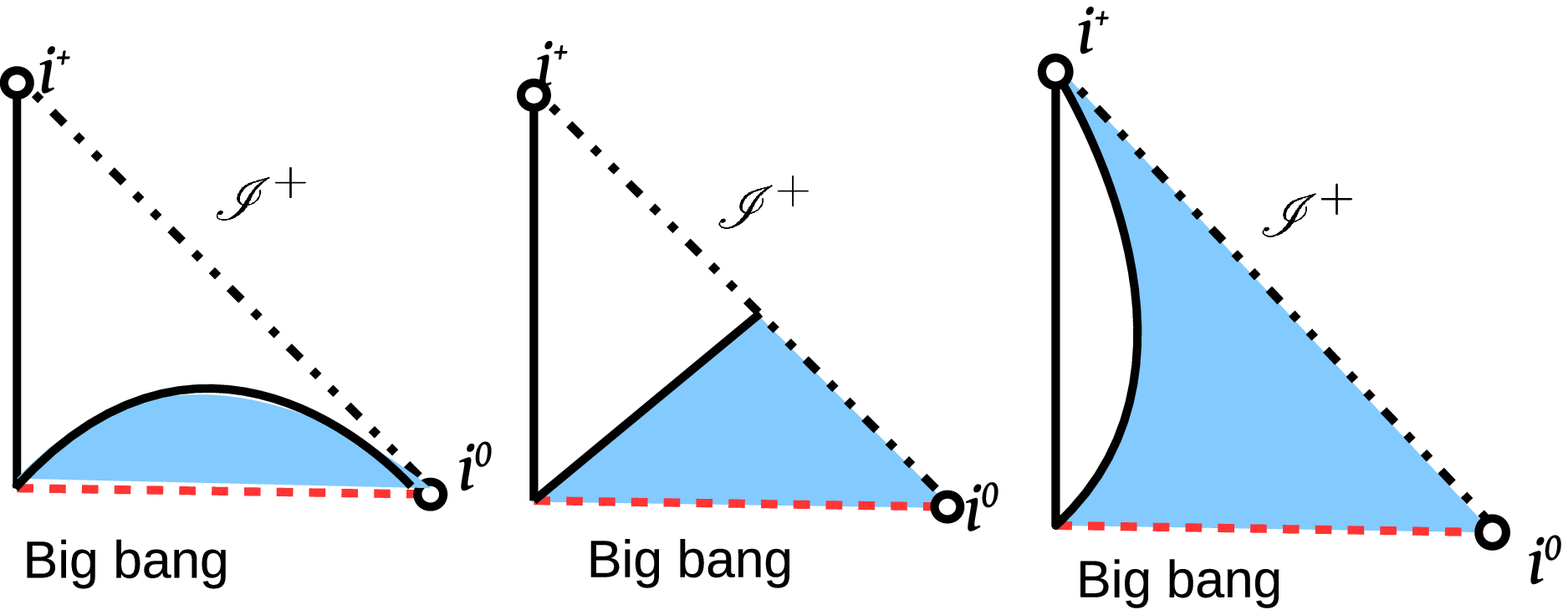}}
\hspace{20mm}
 \subfigure[\label{fg:flatwm13} F2: $w=-1/3$]{\includegraphics[height=0.2\textheight]{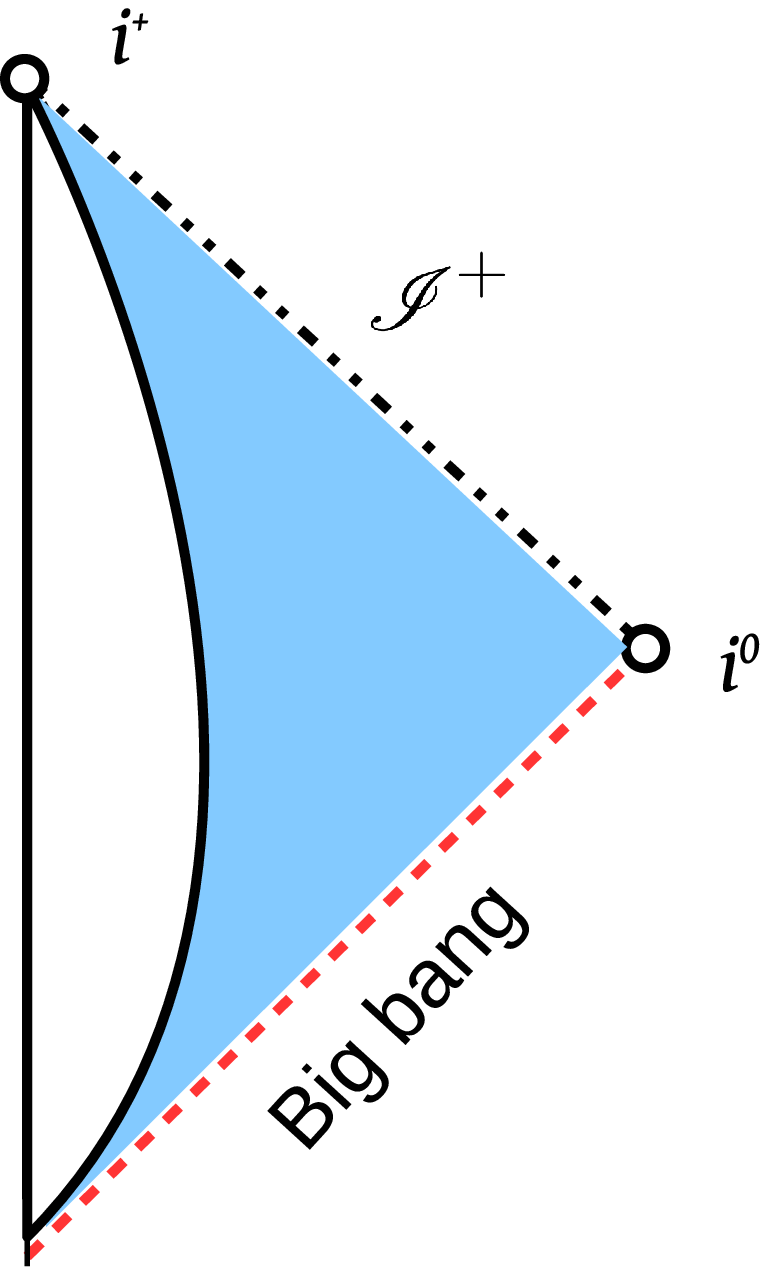}}
\end{center}
\begin{center}
\subfigure[\label{fg:flatwgm1lm13} F3: $-1<w<-1/3$]{\includegraphics[height=0.2\textheight]{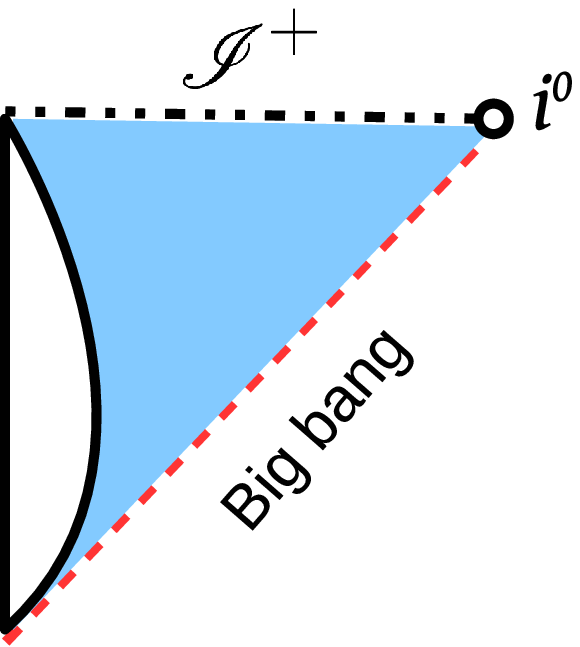}}
\hspace{10mm}
\subfigure[\label{fg:flatwlm1} F4a: $-5/3<w< -1$, F4b:
 $w=-5/3$, F4c: $w<-5/3$]{\includegraphics[height=0.2\textheight]{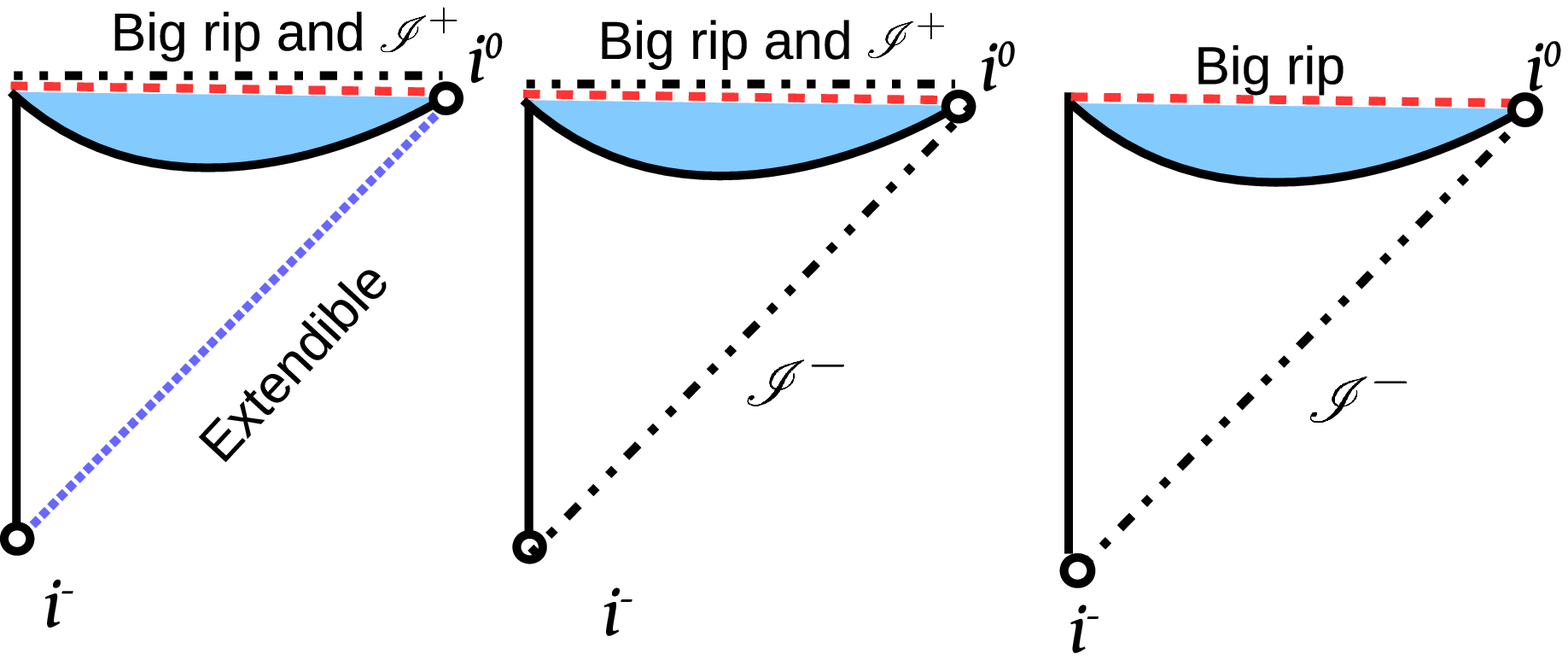}}
\caption{\label{fg:flat} 
The conformal diagrams for the flat FLRW
 solutions. The red dashed lines denote spacetime
 singularities,
while the blue short-dashed line denotes a regular null hypersurfaces at finite
 affine distance.
In Fig.~\ref{fg:flatwgm13}, the left, middle and right panels
correspond to $w>1/3$, $w=1/3$ and $-1/3<w<1/3$,
 respectively, all of which 
are classified as F1.
Figures~\ref{fg:flatwm13} and \ref{fg:flatwgm1lm13} 
correspond to F2 ($w=-1/3$) and F3 ($-1<w<-1/3$), respectively.
In Fig.~\ref{fg:flatwlm1}, the left, middle and right panels
correspond to F4a ($-5/3<w<-1$), F4b ($w=-5/3$) and F4c ($w< -5/3$), respectively.
For $-5/3\le w<-1$, the future big-rip singularity can be reached 
only in an infinite affine length along null geodesics. In this sense,
the future boundary of the spacetime is both a future big-rip singularity
and future null infinity.
This is not the case for $w<-5/3$, where null geodesics also terminate 
at the future big-rip singularity in a finite affine length.
The boundary $t=-r=-\infty$, where $a=0$, 
is a regular null hypersurface at a finite affine length for
 $-5/3<w<-1$,
beyond which the spacetime is at least $C^{2}$-extendible, 
while it is past null infinity for $w\le -5/3$.
}
\end{center}
\end{figure}

The affine parameter along null geodesics
is 
\begin{equation}
 \lambda=\left\{\begin{array}{cc}
  |\eta|^{\frac{5+3w}{1+3w}} & (w\ne -1/3, -1, -5/3)\\
  \ln|\eta| & (w=-5/3) \\
	 \end{array}
\right.
\end{equation}
up to an affine transformation,
so the affine length along null geodesics 
to the spacetime boundaries 
also depends on the value of $w$.
We discuss the different cases 
below,
summarise the spacetime boundaries in Table~\ref{table:flat}
and show the conformal diagrams in Fig.~\ref{fg:flat}.

\begin{itemize}
 \item F1: $w>-1/3$\\ 
In this case, the domain of $\eta$ is 
$0<\eta<\infty$, so
we have the upper half of the conformal diagram 
of Minkowski spacetime.
$\eta=0$ corresponds to a big-bang singularity,
$\eta=r=\infty$ 
to future null infinity and $\eta=\infty$ with $r<\infty$ 
to timelike infinity.
The trapping horizon starts at $\eta=r=0$ and is 
timelike, null and spacelike for $-1/3<w<1/3$, $w=1/3$  
and $w>1/3$, respectively.
There is a particle horizon but no cosmological event horizon.

\item F2: $w=-1/3$\\
In this case, Eq.~(\ref{eq:Friedmann3}) with $K=0$ implies 
$
 a=\tilde{a}_{c}^{1/2}t
$
for $0<t<\infty$, so 
the expansion speed is constant. The conformal time is 
$
 \eta=\tilde{a}_{c}^{-1/2}\ln t
$
and the domain of $\eta$ is 
$-\infty<\eta<\infty$, 
so the conformal diagram is the same as that of Minkowski spacetime. 
$\eta=-\infty$ corresponds to a big-bang singularity, this being
null. 
The Misner-Sharp mass is given by
\begin{equation}
 \frac{2m}{R}=\tilde{a}_{c}r^{2}, 
\end{equation}
so there is a trapping horizon at 
$
 r=\tilde{a}_{c}^{-1/2}
$
and 
this  is timelike and does not cross $r=0$.
The region $r>\tilde{a}_{c}^{-1/2}$ is past trapped.
There is no particle horizon or cosmological event horizon.
The affine parameter along null geodesics is 
\begin{equation}
   \lambda=e^{2\tilde{a}_{c}^{1/2}\eta} 
\end{equation}
up to an affine transformation.
Thus the boundary $\eta=r=\infty$ corresponds to null infinity, 
while the boundary $\eta=-r=-\infty$ is at a finite affine length.

\item F3: $-1<w<-1/3$\\
In this case, the domain of $\eta$ is 
 $-\infty<\eta<0$,
so we have the lower half of the conformal diagram of 
Minkowski spacetime. $\eta=0$ corresponds to future null infinity,
which is spacelike, and $\eta=-\infty$ 
to the big-bang singularity, which  
is null. 
The trapping horizon starts at $\eta=r=0$ and is timelike.
There is a cosmological event horizon but no particle horizon.

\item F4: $w<-1$\\
In this case, $-\infty<t<0$ is mapped to $-\infty<\eta<0$,
so we again have the lower half of the conformal diagram of 
Minkowski spacetime. $\eta=0$ corresponds to the future big-rip singularity,
which is spacelike, and 
$\eta=-\infty$ 
corresponds to $t= -\infty$. 
In the latter limit, there is no divergence in the curvature invariants,
so $\eta=-\infty$ with $r<\infty$ corresponds to past timelike infinity.
Therefore the 3-space emerges from $a=0$ without a singularity.
The trapping horizon terminates at $\eta=r=0$ and is spacelike.
There is a cosmological event horizon but no particle
      horizon. There are then three subcases.

\begin{itemize}
 \item F4a: $-5/3<w<-1$\\
The boundaries $\eta=-\infty$ and $\eta=0$
are at finite and infinite affine lengths along null geodesics, respectively.
Therefore the boundary $\eta=-r=-\infty$ is a regular null hypersurface at 
a finite affine length, 
while the boundary $\eta=0$ is 
a future big-rip singularity which can be reached 
in a finite affine length 
along timelike geodesics but not along null geodesics.
Any null geodesic is complete in the future but incomplete in the past.
This can be understood from 
Ref.~\cite{FernandezJambrina:2006hj}.
Although the detailed analysis of the extension  
beyond the regular null hypersurface
$\eta=-r=-\infty$ is beyond the scope of this paper, 
a possible extension may be obtained by pasting the time reverse of the 
solution onto this null hypersurface. This extension is 
at least $C^{2}$ and the conservation law (\ref{eq:energy_equation})
holds across the null
       hypersurface. 
Any null geodesic is complete in both the future and the past
in this extension.

 \item F4b: $w=-5/3$\\
The boundaries $\eta=-\infty$ and $\eta=0$
are both at an infinite 
affine length along null geodesics.
Therefore the boundary $\eta=-r=-\infty$ is past null infinity, 
while the boundary $\eta=0$ is 
a future big-rip singularity which can be reached in a finite affine length 
along timelike geodesics but not along null geodesics.
Any null geodesic is complete both in the future and in the past. 

\item F4c: $w<-5/3$\\
The boundaries $\eta=-\infty$ and $\eta=0$
are at infinite and 
finite affine lengths along null geodesics, respectively.
Therefore the boundary $\eta=-r=-\infty$ is past null infinity, 
while the boundary $\eta=0$ is 
a future big-rip singularity which can be reached in a finite affine length 
both along timelike and 
null geodesics.
Any null geodesic is incomplete in the future but complete in the past.

\end{itemize}

\end{itemize}

\begin{table}[htbp]
\begin{center}
\caption{\label{table:flat} Flat FLRW solutions} 
\begin{tabular}{llcccccc}
\hline\hline
&&~~~$t$~~~&~~~$ \eta$~~~&~~~$a$~~~&~~~$ \rho$~~~&$ r<\infty$~~~&~~~$ r=\infty$
\\
\hline
F1: &$-1/3<w<\infty$&$ \infty$&$ \infty$&$\infty$&$0$&$i^+$&$\mathscr{I}^+$
\\
&&$0$&$0$&$0$&$\infty$&BB~\footnote{big bang}&$i^0$
\\
\hline
F2: &$w=-1/3$&$\infty$&$\infty$&$\infty$&$0$&$i^+$&$\mathscr{I}^+$
\\
&&$0$&$-\infty$&$0$&$\infty$&BB&BB
\\
\hline
F3: &$-1<w<-1/3$&$\infty$&$0$&$\infty$&$0$&$ \mathscr{I}^+$&$i^0$
\\
&$$&$0$&$-\infty$&$0$&$\infty$&BB&BB
\\
\hline
F4a: &$-5/3<w<-1$&$0$&$ 0$&$ \infty$&$ \infty$& FBR\footnote{future big rip} \& $\mathscr{I}^+$&$ i^0$
\\
&&$ -\infty$&$ -\infty$&$0$&$ 0$&$ i^-$& RNHS\footnote{regular null hypersurface}
\\
\hline
F4b: &$w=-5/3$&$0$&$0$&$\infty$&$ \infty$& FBR \& $\mathscr{I}^+$&$i^0$
\\
&&$-\infty$&$ -\infty$&$0$&0&$i^-$&$ \mathscr{I}^-$
\\
\hline
F4c: &$-\infty<w<-5/3$&$0$&$0$&$ \infty$&$ \infty$& FBR &$ i^0$
\\
&&$ -\infty$&$ -\infty$&$0$&$0$&$ i^-$&$ \mathscr{I}^-$
\\
\hline\hline
\end{tabular}

\end{center}
 
\end{table}

\section{Positive-curvature FLRW solutions}
\label{sec:PC}
As can be seen in Eq.~(\ref{eq:Friedmann2}) with $K=1$, for $w>-1/3$
the scale factor begins at zero, increases to $a_{c}$ and then 
decreases to zero. However, for $w<-1/3$,
it begins at infinity, decreases to $a_{c}$ and then 
increases to infinity.
For $w=-1/3$, the expansion speed 
is constant.
More precisely, for $w\ne -1/3$, 
the solution of Eq.~(\ref{eq:Friedmann_dust}) with $K=1$ is 
\begin{eqnarray}
 \tilde{a}&=&\tilde{a}_{c} \frac{1-\cos\tilde{\eta}}{2} \, ,~~\tilde{t}=\tilde{a}_{c} \frac{\tilde{\eta}-\sin\tilde{\eta}}{2} \, ,
\end{eqnarray}
where $0<\tilde{\eta}<2\pi$.
The form of $a(t)$ is shown in Fig.~\ref{scalefactor}.
The metric is conformal to that of the 
Einstein static universe with the conformal time being
\begin{equation}
 \eta=\frac{1}{1+3w}\tilde{\eta} \, . 
\end{equation}
A straightforward calculation 
gives
\begin{equation}
 \frac{2m}{R}=\frac{\sin^{2}r}{\sin^{2}(\tilde{\eta}/2)} \, ,
\end{equation}
so there are two trapping horizons at
\begin{equation}
 \eta=\frac{2}{1+3w}r \, , \quad  \frac{2}{1+3w}(\pi-r)
\end{equation}
and these cross 
each other at $(r,\eta)=(\pi/2,\pi/(1+3w))$. We note that a
photon can circumnavigate the universe before it 
crunches for $-1/3<w <0$.

As for the affine lengths of null geodesics, we only have
      to focus on $\eta=0$ and $2\pi/(1+3w)$ for $w\ne -1/3, -1$.
In this case, as $\eta\to 0$,
      we find 
\begin{equation}
 \lambda\simeq \left\{\begin{array}{cc}
  |\eta|^{\frac{5+3w}{1+3w}} & (w\ne -1/3, -1, -5/3)\\
  \ln|\eta| & (w=-5/3) \\
	 \end{array}
\right.
\label{eq:null_affine_pc}
\end{equation}
up to an affine transformation. The affine parameter in the limit
$\eta\to 2\pi/(1+3w)$ corresponding to another boundary is 
given by replacing $\eta$ with $2\pi/(1+3w)-\eta$ in Eq.~(\ref{eq:null_affine_pc}).

\begin{figure}[htbp]
\begin{center}
\subfigure[\label{fg:pcwgm13} P1: $w> 1/3$, $w=1/3$,
 $-1/3<w<1/3$]{\includegraphics[height=0.2\textheight]{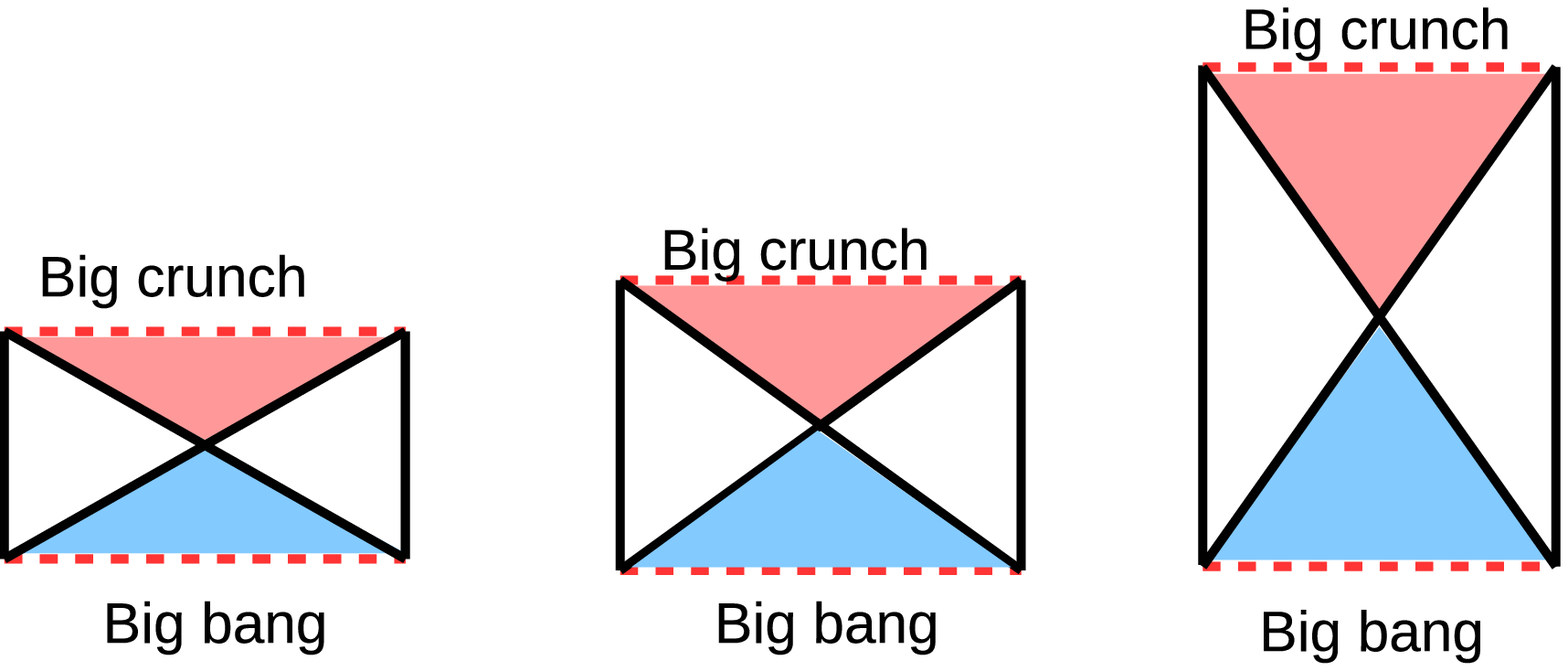}}
\hspace{10mm}
\subfigure[\label{fg:pcwm13} P2: $w=-1/3$]{\includegraphics[height=0.25\textheight]{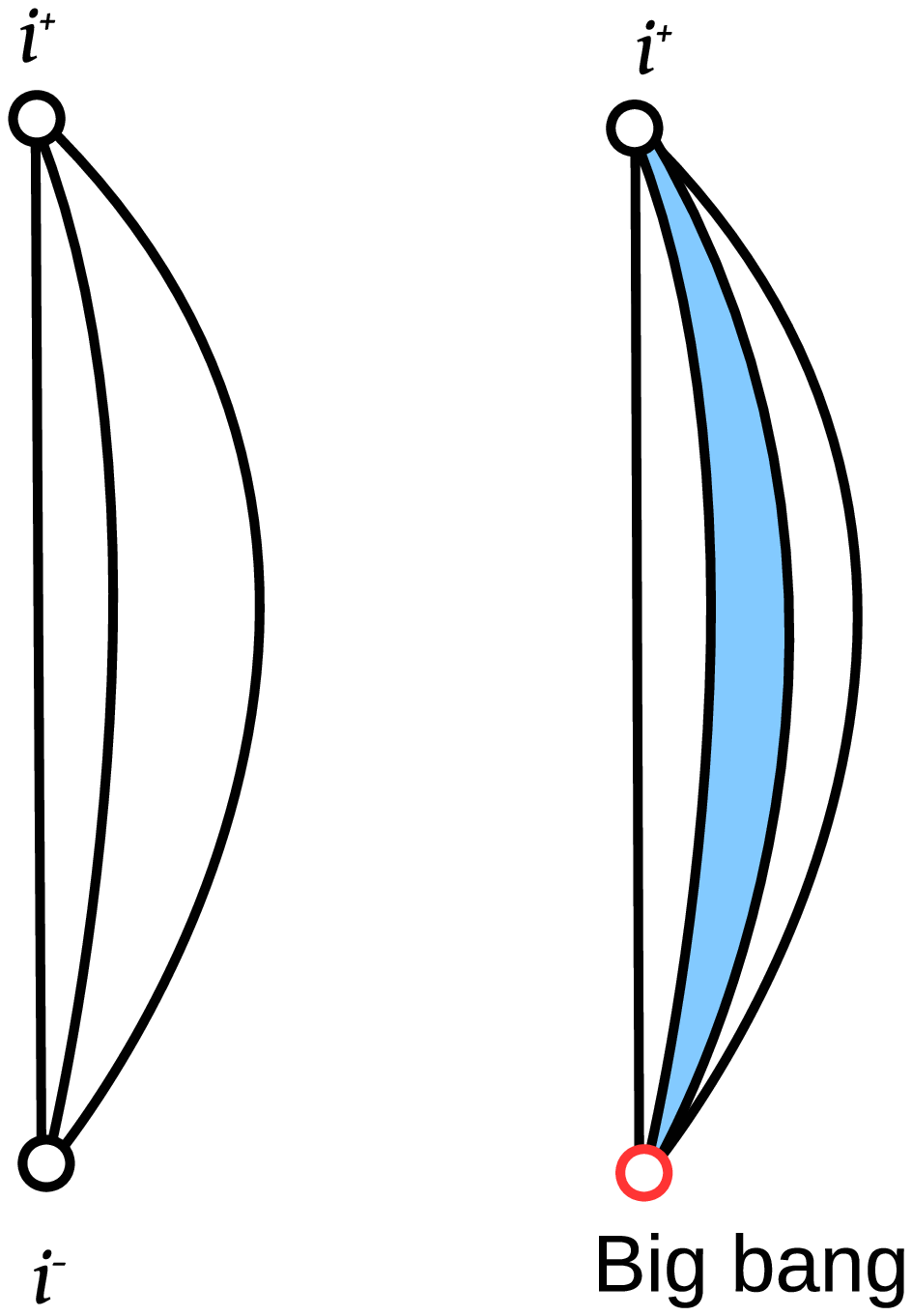}}
\end{center}
\begin{center}
\subfigure[\label{fg:pcwgm1lm13} P3:~$-1<w< -1/3$]{\includegraphics[height=0.3\textheight]{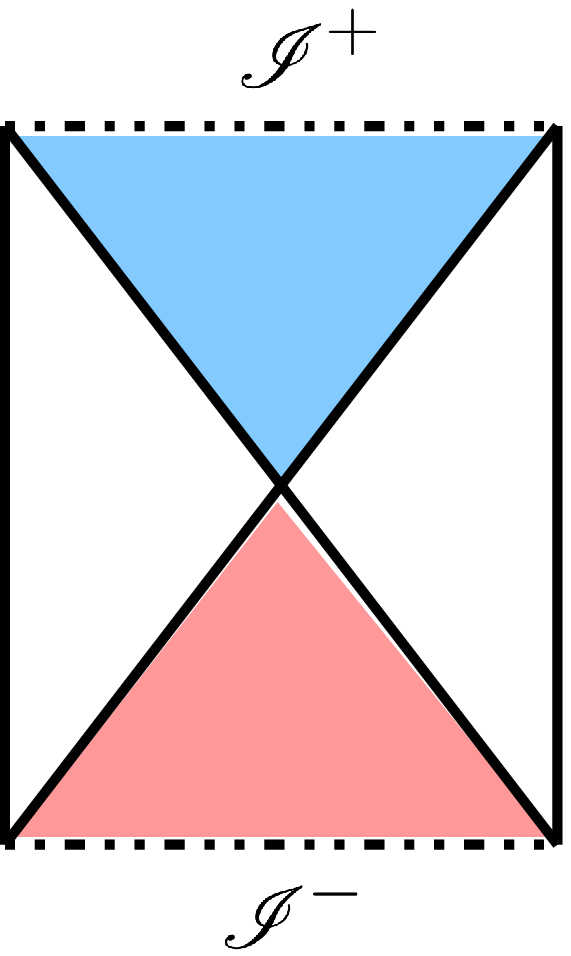}}
\hspace{10mm}
 \subfigure[\label{fg:pcwlm1} P4a: $-5/3\le w< -1$, 
P4b: $w<-5/3$]{\includegraphics[height=0.2\textheight]{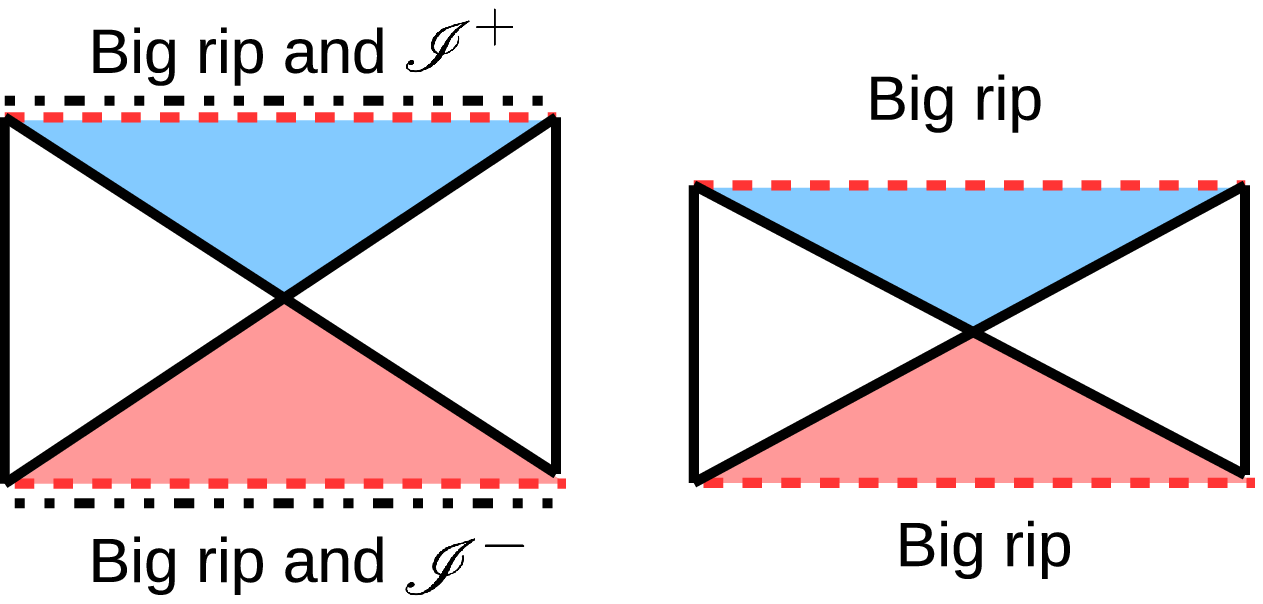}}
\caption{\label{fg:positive_curvature} 
The conformal diagrams for the positive-curvature FLRW
 solutions. In Fig.~\ref{fg:pcwgm13}, the left, middle and right panels
correspond to $w>1/3$, $w=1/3$ and $-1/3<w<1/3$, respectively, all of
 which are classified as P1.
In Fig.~\ref{fg:pcwm13}, the left and right panels 
correspond to $w=-1/3$ with $\tilde{a}_{c}=1$ and $\tilde{a}_{c}>1$,
respectively, both of which are classified as P2.
Figure~\ref{fg:pcwgm1lm13} corresponds to P3 ($-1<w<-1/3$).
In Fig.~\ref{fg:pcwlm1}, the left and right panels
correspond to P4a ($-5/3\le w<-1$) and P4b ($w<-5/3$), respectively.
}
\end{center}
\end{figure}

Table~\ref{table:positive_curvature} summarises 
the spacetime boundaries and 
Fig.~\ref{fg:positive_curvature} shows the 
conformal diagrams for $K=1$. We now discuss the various cases.
Note that there are both past and future trapped regions 
and
also both particle and cosmological event horizons in this case.
\begin{itemize}
 \item P1: $w>-1/3$\\
In this case, the domain of $\eta$ is $0<\eta<2\pi/(1+3w)$.
The scale factor is always smaller than or equal to
$a_{c}$ and $\rho$ is a decreasing function of $a$.
Therefore $\rho$ is bounded from below by its value 
at maximum expansion, which occurs at $\eta=\pi/(1+3w)$.
Thus $\eta=0$ and $\eta=2\pi/(1+3w)$ correspond to big-bang and 
big-crunch singularities, respectively. 
The trapping horizons are spacelike for $w>1/3$, 
null for $w=1/3$ and timelike for $-1/3<w<1/3$.
Because of the recollapsing dynamics, 
the region with
$0<\eta<\pi/(1+3w)$ and
       $[(1+3w)/2]\eta<r<\pi-[(1+3w)/2]\eta$ is past trapped, 
while the one with
$\pi/(1+3w)<\eta<2\pi/(1+3w)$ and
       $\pi-[(1+3w)/2]\eta<r<[(1+3w)/2]\eta$ is future trapped. 

\item P2: $w=-1/3$\\
In this case, 
Eq.~(\ref{eq:Friedmann3}) gives 
two solutions. If $\tilde{a}_{c}=1$, 
then $a=a_{0}$ and $\eta=t/a_{0}$, where $a_{0}$ is a constant of
      integration. The spacetime is then identical to the Einstein
static universe, with
no singularity, and the domain of $\eta$
is $-\infty<\eta<\infty$.
If $\tilde{a}_{c}>1$ and the 
expanding branch is chosen, $a= \sqrt{\tilde{a}_{c}-1} \, t $.  
The collapsing branch is just its time reverse.
The conformal time is given by
$
 \eta=({1}/{\sqrt{\tilde{a}_{c}-1}})\ln t \, 
$
and the domain of $\eta$ being  $-\infty<\eta<\infty$.
In both cases, 
the $(\eta, r)$ 
part of the metric is
already 
conformally flat. However, 
the domain of $(\eta,r)$, which is 
$-\infty<\eta<\infty$ and $0\le r\le \pi$, is unbounded.
By the transformation
\begin{eqnarray}
 2\eta=\tan \left( \frac{T+X}{2} \right)+\tan \left(\frac{T-X}{2} \right), ~~ 
 2r=\tan  \left(\frac{T+X}{2} \right) -\tan  \left( \frac{T-X}{2} \right) \, ,
\end{eqnarray}
we then obtain
the following form for the metric:
\begin{eqnarray}
 ds^{2}&=&\frac{1}{4}a^{2}(\eta)\sec^{2}  
\left( \frac{T+X}{2} \right) \sec^{2}  \left( \frac{T-X}{2}
					     \right)
\nonumber \\ 
&\times& \left[-dT^{2}+dX^{2}
+4\cos^{2}  \left( \frac{T+X}{2} \right) \cos^{2}  \left( \frac{T-X}{2} \right) 
\sin^{2}r \, d\Omega^{2}\right],
\end{eqnarray}
where $\eta$ and $r$ are
regarded as functions of $T$ and $X$ and 
the domain of $(X,T)$ is 
\begin{eqnarray}
&&-\pi<T-X<\pi \, ,~~ -\pi<T+X<\pi \, , ~~X\ge 0, \, ~~\nonumber \\
&&\tan  \left(\frac{T+X}{2} \right) -\tan \left(\frac{T-X}{2} \right) \le  2\pi \,.
\end{eqnarray}
$\eta=-\infty$ is transformed to $(X,T)=(0,-\pi)$, which is past timelike
infinity for $\tilde{a}_{c}=1$ but
a big-bang 
singularity for $\tilde{a}_{c}>1$.
The timelike curve $r=0$ is 
transformed to $X=0$, while $r=\pi$ is transformed to 
\begin{equation}
\tan  \left( \frac{T+X}{2} \right) -\tan  \left(\frac{T-X}{2} \right)= 2\pi \, . 
\end{equation}
This 
can be written as
\begin{equation}
 T-X=2\arctan \left[\tan  \left( \frac{T+X}{2} \right)-2\pi\right],
\end{equation}
which 
passes through  $(X,T)=(2\arctan\pi,0)$, 
	 $(0,\pi)$ and $(0,-\pi)$. 
In this case,
\begin{equation}
 \frac{2m}{R}=\tilde{a}_{c}\sin^{2}r \, , 
\end{equation}
so  there is a trapping horizon at $r=\pi/2$, which 
is timelike, but no trapped region for $\tilde{a}_{c}=1$.
For $\tilde{a}_{c}>1$, there are two trapping horizons at 
\begin{equation}
 r=\arcsin (\tilde{a}_{c}^{-1/2}),~~\pi-\arcsin (\tilde{a}_{c}^{-1/2}) \, . 
\end{equation}
These are both timelike. The region with
$\arcsin(\tilde{a}_{c}^{-1/2})<r<\pi-\arcsin(\tilde{a}_{c}^{-1/2})$ is past trapped.
For $w=-1/3$, the affine parameter is 
\begin{eqnarray}
 \lambda&=& \left\{\begin{array}{cc}
   \eta & (\tilde{a}_{c}=1) \\
   e^{2\sqrt{\tilde{a}_{c}-1}\eta} & (\tilde{a}_{c}>1)
		 \end{array}\right.
\end{eqnarray}
up to an affine transformation. Thus, for $\tilde{a}_{c}=1$, 
both the boundaries $\eta=\pm \infty$ are at an infinite affine length 
along null geodesics.
For $\tilde{a}_{c}>1$, 
 the boundaries $\eta=\infty$ and $\eta=-\infty$ are at infinite and finite affine lengths along 
      null geodesics, respectively.

\item P3: $-1<w<-1/3$\\
In this case, the scale factor is never less than
$a_{c}$ and $\rho$ is a
decreasing function of $a$. 
Therefore $\rho$ is bounded from above by the value at the bounce,
 which occurs at $\eta=\pi/(1+3w)$. 
The domain of $\eta$ is 
$2\pi/(1+3w)<\eta<0$, 
with $\eta=0$ and
$2\pi/(1+3w)$ corresponding to null infinities rather than singularities. 
The trapping horizons are timelike. 
Because of the bouncing dynamics, 
the region with
$\pi/(1+3w)<\eta<0$ and
       $[(1+3w)/2]\eta<r<\pi-[(1+3w)/2]\eta$ is past trapped, 
while that with
 $2\pi/(1+3w)<\eta<\pi/(1+3w)$ and
       $\pi-[(1+3w)/2]\eta<r<[(1+3w)/2]\eta$ is future trapped. 

\item P4: $w<-1$\\
In this case, the scale factor is never less than
$a_{c}$ and the 
energy density is an increasing function of $a$. 
The domain of $\eta$ is 
$2\pi/(1+3w)<\eta<0$, 
with $\eta=2\pi/(1+w)$ and
$0$ corresponding to past and future big-rip singularities, respectively.
The bounce occurs at $\eta=\pi/(1+3w)$, where $\rho$
      reaches its minimum.
The trapping horizons are spacelike.
Because of the bouncing dynamics, 
the region with
$\pi/(1+3w)<\eta<0$ and
       $[(1+3w)/2]\eta<r<\pi-[(1+3w)/2]\eta$ is past trapped, 
while that with
$2\pi/(1+3w)<\eta<\pi/(1+3w)$ and
       $\pi-[(1+3w)/2]\eta<r<[(1+3w)/2]\eta$ is future trapped. 
There are then two subcases.

\begin{itemize}
 \item P4a: $-5/3\le w<-1$\\
The boundaries $\eta=2\pi/(1+3w)$ and $\eta=0$ are both null infinities.
Therefore they are both 
big-rip singularities and 
null infinities, simultaneously.
 \item P4b: $w<-5/3$\\
The boundaries $\eta=2\pi/(1+3w)$ and $\eta=0$ are both at a finite affine
       length. 
Therefore they are 
big-rip singularities but not 
null infinities.
\end{itemize}
\end{itemize}

\begin{table}[htbp]
\begin{center}
\caption{\label{table:positive_curvature} Positive-curvature FLRW
 solutions}
\begin{tabular}{llcccccccc}
\hline\hline
&&~~~$t$~~~&~~~$\tilde t$~~~~~&~~~~~~~$\eta$~~~~~~~&~~~~$\tilde \eta$~~~~~&~~~$a$~~~&~~~$ \tilde a$~~~&~~~$\rho$~~~&~~~$r<\infty$
\\
\hline
P1: &$-1/3<w<\infty$&$t_0$& $\pi \tilde{a}_{c}$&$2\pi/(1+3w)$&$2\pi$&$0$&$0$&$\infty$&BC~\footnote{big crunch}
\\
&&$0$&$0$&$0$&$0$&$0$&$0$&$\infty$&BB~\footnote{big bang}
\\
\hline
P2: &$w=-1/3$ ($\tilde a_c=1$)&$\infty$&--&$\infty$&--&$ a_0$&--&$ \rho_0$&$ i^+$
\\
&&$-\infty$&--&$\infty$&--&$ a_0$&--&$\rho_0$&$ i^-$
\\
\hline
P2: &$w=-1/3$ ($\tilde a_c>1$)&$\infty$&--&$\infty$&--&$ \infty$&--&$0$&$ i^+$
\\
&&$0$&--&$ -\infty$&--&$0$&--&$\infty$&BB
\\
\hline
P3: &$-1<w<-1/3$&$ \infty$&$0$&$0$&$0$&$\infty$&$0$&$ 0$&$\mathscr{I}^+$
\\
&&$ -\infty$&$\pi \tilde a_c$&$2\pi/(1+3w)$&$2\pi$&$\infty$&$0$&$0$&$\mathscr{I}^-$
\\
\hline
P4a: &$-5/3\leq w<-1$&$t_0$&$0$&$0$&$0$&$\infty$&$0$&$\infty$&FBR~\footnote{future big
 rip} \& $\mathscr{I}^+$
\\
&$$&$0$&$\pi \tilde a_c$&$2\pi/(1+3w)$&$2\pi$&$\infty$&$0$&$\infty$&PBR~\footnote{past big rip} \& $\mathscr{I}^-$
\\
\hline
P4b: &$-\infty<w<-5/3$&$t_0$&$0$&$0$&$0$&$\infty$&$0$&$\infty$&FBR
\\
&$$&$0$&$\pi \tilde a_c$&$2\pi/(1+3 w)$&$2\pi$&$\infty$&$0$&$\infty$&PBR
\\
\hline\hline
\end{tabular} 
\end{center}
\end{table}

\section{Negative-curvature FLRW solutions}
\label{sec:NC}

For $K=-1$, the conformally static form of the 
metric 
(\ref{eq:FLRW_metric'})
can be written as~\cite{Sato:1996} 
\begin{equation}
 ds^{2}=a^{2}(\eta)\frac{-d\eta'^{2}+dr'^{2}+\sin^{2}r'd\Omega^{2}}{\cos(r'+\eta')\cos(r'-\eta')} \, ,
\label{eq:metric_conformal_to_SE}
\end{equation}
where 
\begin{equation}
 \tan\eta'=\frac{\sinh\eta}{\cosh r} \, ,~~\tan r'=\frac{\sinh r}{\cosh\eta} \, .
\end{equation}
Therefore the metric 
is conformal to
the Einstein static universe. In this case -- and {\it only} this case -- 
the energy density can be
negative,
so we consider this possibility below.

\subsection{Vacuum}
\begin{figure}[htbp]
\begin{center}
\includegraphics[height=0.3\textheight]{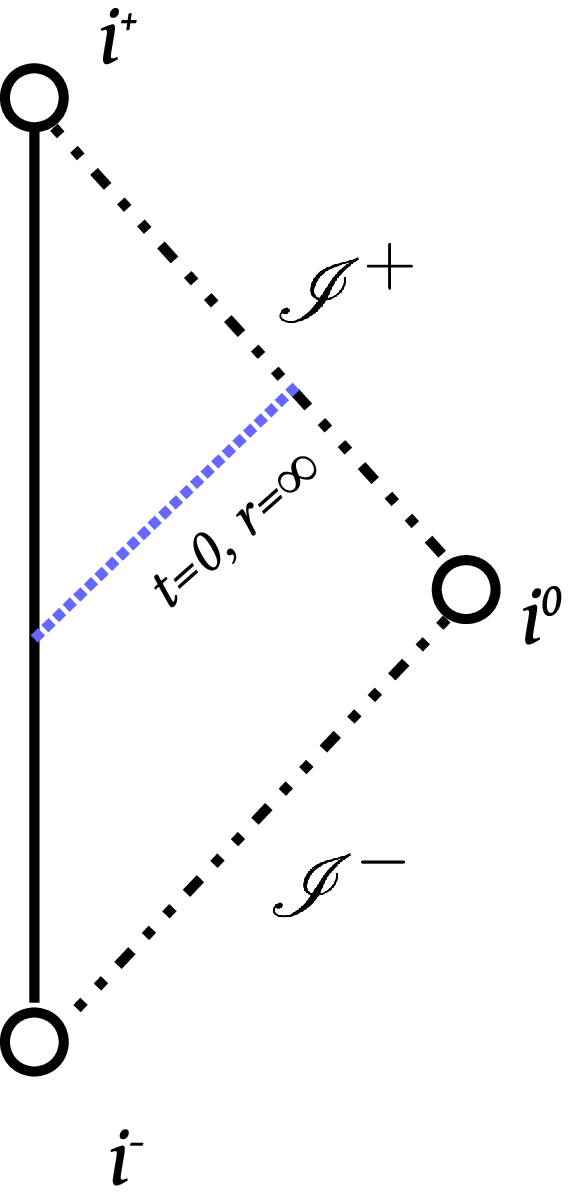}
\end{center}
\caption{\label{fg:Milne} 
The conformal diagram for the negative-curvature FLRW solution
 or the Milne universe is shown as part of the Minkowski spacetime, which is
its maximal extension.}
\end{figure}

For pedagogical completeness, we consider
the vacuum case, in which the solution is 
part of Minkowski spacetime. The 
 Friedmann equation 
with $\rho=0$ and $K=-1$ gives
$a=t$
with $0<t<\infty$ for  the expanding branch,
corresponding to the Milne universe. The collapsing branch is 
 the time reverse of this.
Since 
$ \eta=\ln t,$
the domain $0<t<\infty$ is mapped to $-\infty<\eta<\infty$. 
In the form which is conformal to the Einstein
static universe, the domain of $\eta'$ and $r'$ is 
the intersection of
$-\pi/2<\eta'<\pi/2$, $0\le r'<\pi/2$, $-\pi/2<r'-\eta'<\pi/2$ and
$-\pi/2<r'+\eta'<\pi/2$. 
The affine parameter along null geodesics is given by
$\lambda=e^{2\eta}$ up to an affine transformation. Therefore, the past
boundary $t=0$ or $\eta= -\infty$ is at a finite affine length along 
null geodesics.
As it is well known, the metric can be transformed to the standard
Minkowski form with the substitution
$
 T=t\cosh r$ and  $X=t\sinh r$.
While the domain of $T$ and $X$ is 
originally
the intersection of $0<T<\infty$, $0\le X<\infty$ and $T>
X$, it can be maximally extended beyond $T=X$ to the region
$-\infty<T<\infty$ and $0\le X<\infty$. 
The past boundaries $\{t=0, \,
r<\infty\}$ and $\{t=0, \,
r=\infty\}$ of the Milne universe
are transformed to the point $T=X=0$ and the null hypersurface
$T=X$,
respectively. 
We will see below that this extension is also possible for $\rho\ne 0$
and $w<-1$.
Table~\ref{table:Milne} summarises the spacetime boundaries and 
Fig.~\ref{fg:Milne} shows the conformal diagram of the Milne universe.

\begin{table}[htbp]
\begin{center}
\caption{\label{table:Milne} Negative-curvature vacuum FLRW
 solution or the Milne universe}
\begin{tabular}{lcccccc}
\hline\hline
&~~~$t$~~~&~~~$\eta$~~~&~~~$a$~~~&~~~$ \rho$~~~&~~~$ r<\infty$~~~&~~~$ r=\infty$
\\
\hline
Milne &$ \infty$&$\infty$&$\infty$&$0$&$i^+$&$ \mathscr{I}^+$
\\
&$0$&$-\infty$&$0$&$0$&regular &RNHS~\footnote{regular null hypersurface}
\\
\hline\hline
\end{tabular}
\end{center}
\end{table}

\subsection{Positive energy density}
If $\rho$ is positive, Eq.~(\ref{eq:Friedmann}) shows that the scale factor $a$ is a monotonic 
function of $t$.
For the expanding solution, it  begins at $0$ at a finite value of $t$ and then increases to $\infty$
as $t$ increases. More precisely, 
for $w\ne -1/3$, 
the expanding
solution of Eq.~(\ref{eq:Friedmann_dust}) with $K=-1$ 
is given by
\begin{eqnarray}
 \tilde{a}&=&\tilde{a}_{c} \frac{\cosh\tilde{\eta}-1}{2} \, ,~~\tilde{t}=\tilde{a}_{c} \frac{\sinh\tilde{\eta}-\tilde{\eta}}{2} \, , 
\label{eq:tilde_variables_NP}
\end{eqnarray}
where $0<\tilde{\eta}<\infty$ 
and $\tilde{\eta}$ is related to $\eta$ by 
\begin{equation}
 \eta=\frac{1}{1+3w}\tilde{\eta} \, .
\end{equation}
The collapsing
solution is the time reverse of this. 
The form of $a(t)$ is shown in Fig.~\ref{scalefactor}.
The Misner-Sharp mass is given by 
\begin{equation}
 \frac{2m}{R}=\frac{\sinh^{2}r}{\sinh^{2}(\tilde{\eta}/{2})} \, ,
\end{equation}
so there is a trapping horizon at 
\begin{equation}
 \eta=\frac{2}{1+3w}r \, .
\end{equation}
This  is spacelike
for $w>1/3$ and $w<-1$, null for $w=1/3$ and timelike for $-1/3<w<1/3$.
The region $r>|(1+3w)\eta|/2$ is past trapped.

\begin{figure}[htbp]
\begin{center}
\subfigure[\label{fg:ncwgm13} NP1: $w> 1/3$, $w=1/3$,
 $-1/3<w<1/3$]{\includegraphics[height=0.2\textheight]{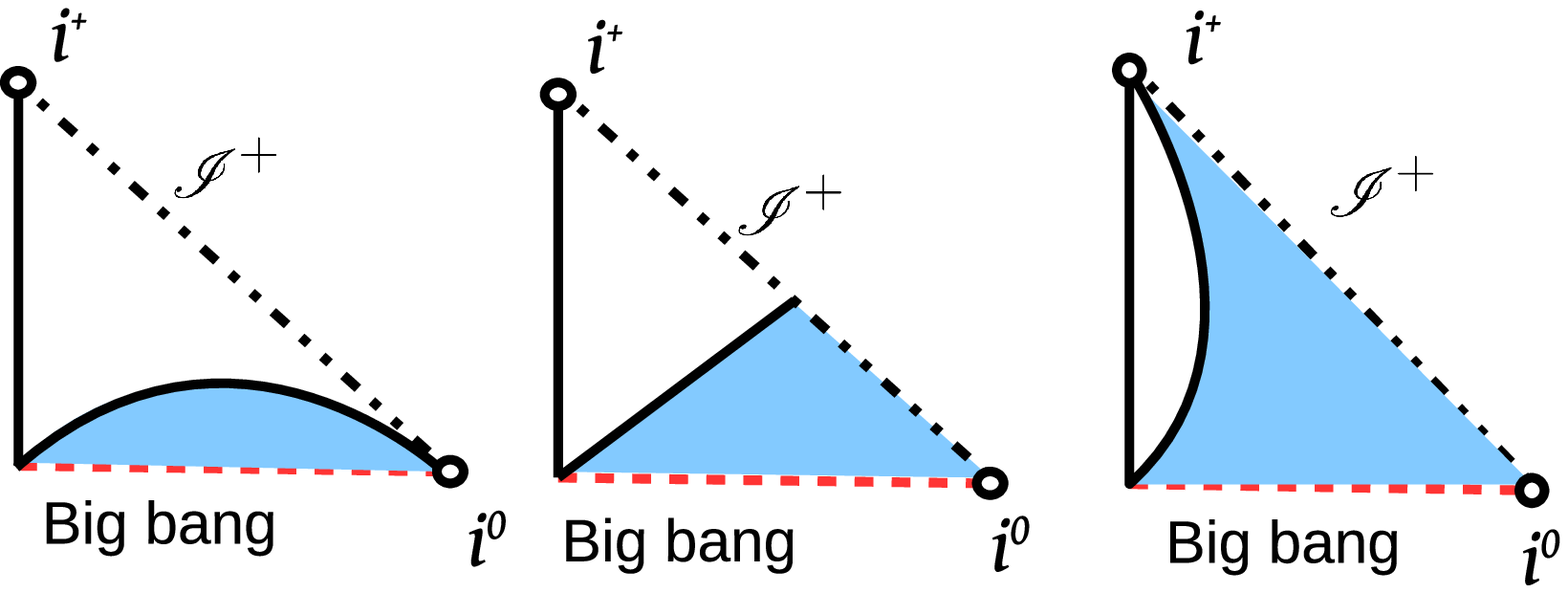}}
\hspace{10mm}
 \subfigure[\label{fg:ncwm13} NP2: $w=-1/3$]{\includegraphics[height=0.25\textheight]{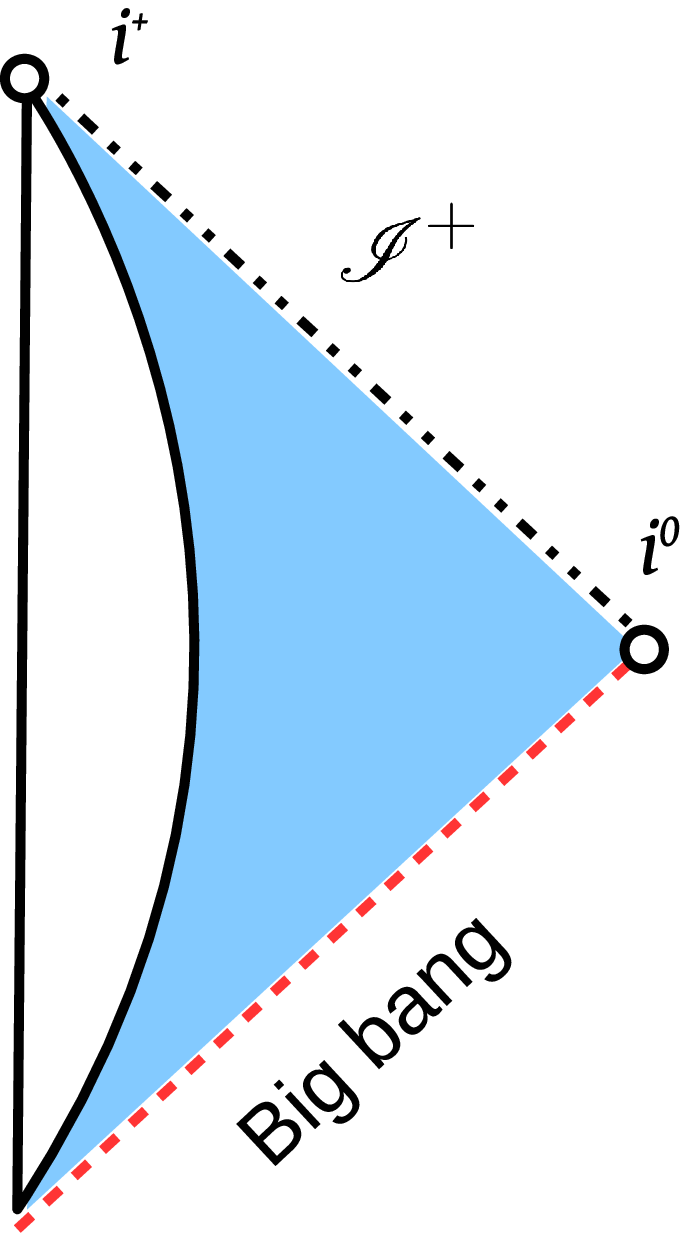}}
\end{center}
\begin{center}
\subfigure[\label{fg:ncwgm1lm13} NP3: $-1<w<-1/3$]{\includegraphics[height=0.2\textheight]{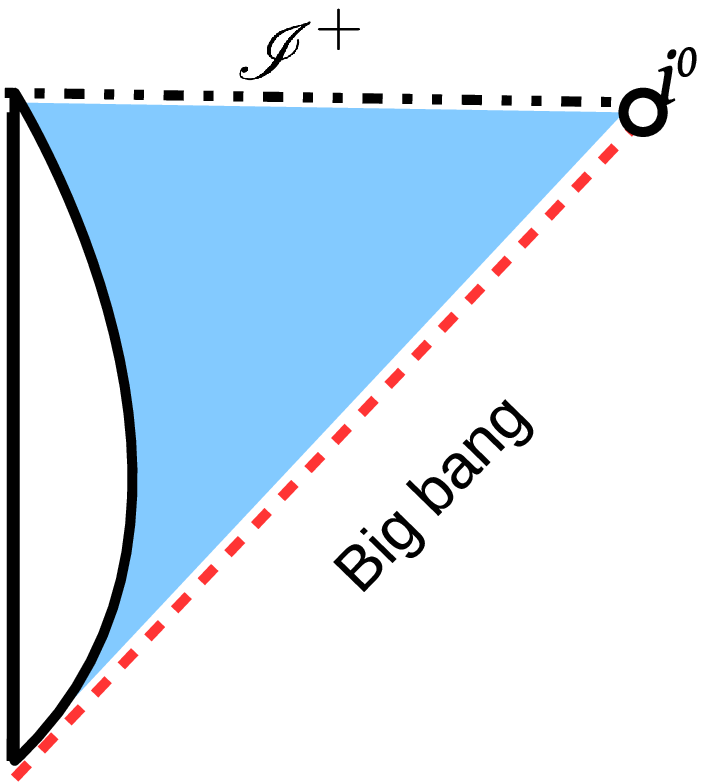}}
\hspace{10mm}
\subfigure[\label{fg:ncwlm1} NP4a: $-5/3\le w< -1$, NP4b: $w<-5/3$]{\includegraphics[height=0.2\textheight]{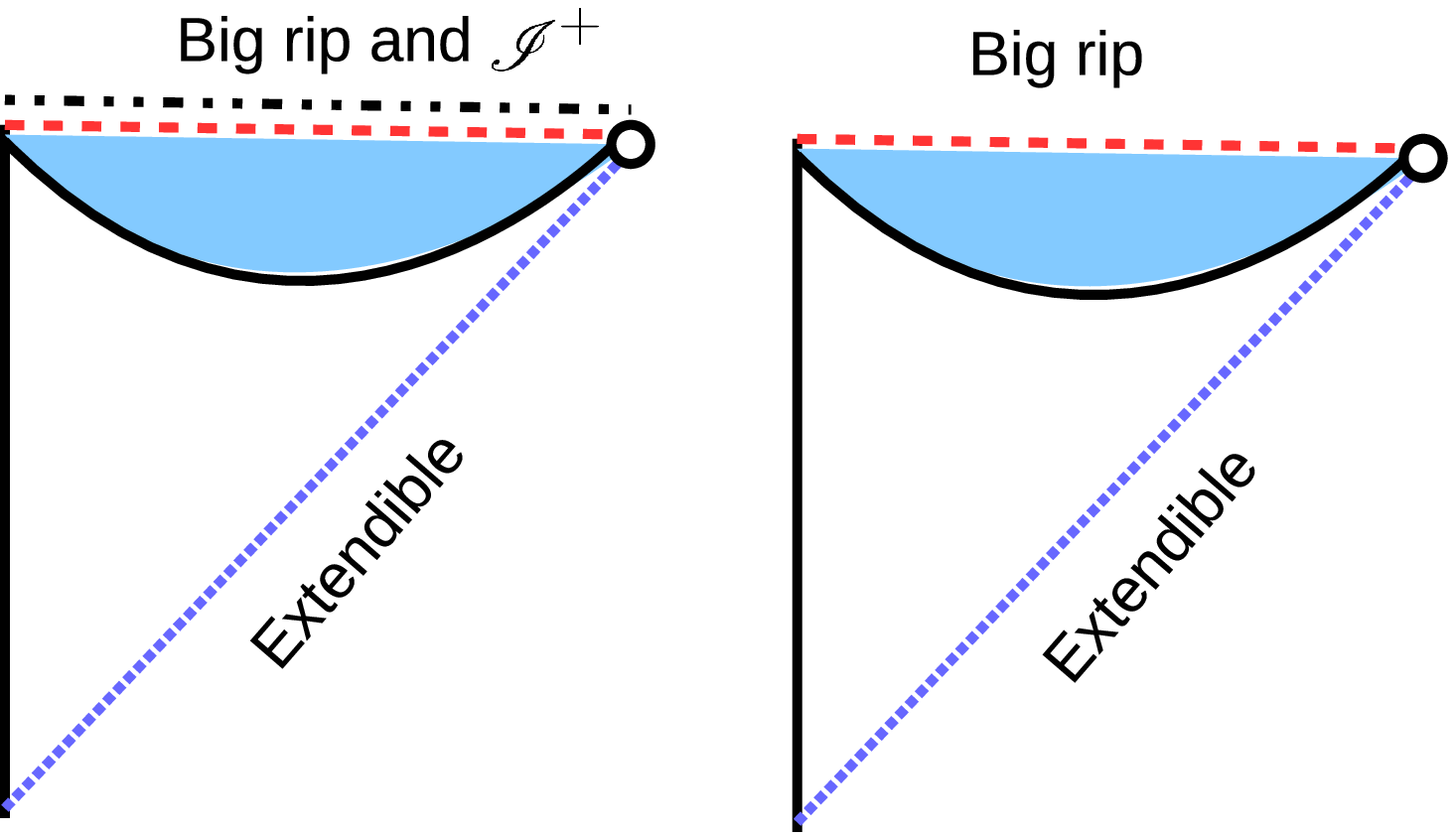}}
\caption{\label{fg:negative_curvature} 
The conformal diagrams for the negative-curvature FLRW
 solutions with a positive energy density. 
In Fig.~\ref{fg:ncwgm13}, the left, middle and right panels
correspond to $w>1/3$, $w=1/3$ and $-1/3<w<1/3$, respectively, 
all of which are classified as NP1.
Figures~\ref{fg:ncwm13} and \ref{fg:ncwgm1lm13} correspond to
NP2 ($w=-1/3$) and NP3 ($-1<w<-1/3$), respectively.
In Fig.~\ref{fg:ncwlm1}, the left and right panels correspond to
NP4a ($-5/3\le w<-1$) and NP4b ($w<-5/3$), respectively.
The blue short-dashed lines for $w<-1$ are  
regular null hypersurfaces at finite affine lengths, beyond 
which the spacetime is at least $C^{2}$-extendible.}
\end{center}
\end{figure}

As for the affine lengths of null geodesics, 
for $\eta\to 0$ we find the behaviour of the affine parameter is the
same as in the positive curvature case
and given by  
Eq.~(\ref{eq:null_affine_pc})
up to an affine transformation. 
For $w\ne -1/3$, the 
affine parameter along null geodesics 
for $\eta \to \pm\infty $ is as
follows:
\begin{equation}
 \lambda\simeq \left\{\begin{array}{cc}
  e^{2|\eta|} & (w>-1/3)\\
  e^{-2|\eta|} & (w<-1/3) \\
	 \end{array}
\right.
\label{eq:nc_null_affine}
\end{equation}
up to an affine transformation. 
Table~\ref{table:negative_curvature} summarises the 
spacetime boundaries and 
Fig.~\ref{fg:negative_curvature} shows the 
conformal diagrams for $K=-1$ with a positive energy density.
Figure~\ref{fg:negative_curvature} looks almost identical to
Fig.~\ref{fg:flat} but is very different in the past boundaries 
for $w<-1$. We now discuss the various cases. 
\begin{itemize}
 \item NP1: $w>-1/3$\\
In this case, $\dot{a}\to 1$ as $a\to \infty$, 
corresponding to infinity at $t= \infty$ or $\eta=\infty$ rather than a big-rip singularity.
However, $t=0$ or $\eta=0$ corresponds to a big-bang singularity.
The domain of $\eta$ and $r$ is 
 $0<\eta<\infty$
and $0\le r<\infty$, this being mapped to 
$0<\eta'<\pi/2$, $0\le r'< \pi/2$ and $\eta'+r'<\pi/2$.
The big-bang singularity at $\eta=0$ is mapped 
to $\eta'=0$. The boundary $\eta=\infty$ with $r<\infty$ is mapped to
$(r',\eta')=(0,\pi/2)$, corresponding to future timelike infinity.
The boundary $r=\infty$ with $\eta<\infty$ is mapped to $(r',\eta')=(\pi/2,0)$,
corresponding to spatial infinity.
The boundary $\eta=r=\infty$ is mapped to $\eta'+r'=\pi/2$, corresponding 
to future null infinity. 
The trapping horizon starts at $\eta'=r'=0$.
There is a particle horizon but no cosmological event horizon.

\item NP2: $w=-1/3$\\
In this case,
the Friedmann equation gives
$
 a=\sqrt{\tilde{a}_{c}+1} \, t,
$
where $0<t<\infty$ and the expanding branch is chosen.
The collapsing one is the time reverse of this. Since
$
 \eta= (1/\sqrt{\tilde{a}_{c}+1})\ln t ,
$
the domain of $\eta$ is 
 $-\infty<\eta<\infty$, where 
$\eta=-\infty$ corresponds to the big-bang singularity.
In this case
\begin{equation}
  \frac{2m}{R}=\tilde{a}_{c}\sinh^{2}r,
\end{equation}
so there is a trapping horizon at 
$
 r=\mbox{arcsinh}(\tilde{a}_{c}^{-1/2} )\, ,
$
which is timelike.
The region $r>\mbox{arcsinh}(\tilde{a}_{c}^{-1/2})$ is past trapped.
There is no particle horizon or cosmological event horizon.
The affine parameter is given by 
\begin{equation}
 \lambda=
   e^{2\sqrt{\tilde{a}_{c}+1}\eta} 
\end{equation}
up to an affine transformation. Thus the 
      boundary $\eta=r=\infty$ corresponds to null infinity, while 
      the boundary $\eta=-r=-\infty$ is at a finite affine length.

\item NP3: $-1<w<-1/3$\\
In this case, as $a\to \infty$, the curvature term becomes subdominant in the 
Friedmann equation and so the solution asymptotically approaches the 
flat model.
The domain of $\eta$ and $r$ is 
$-\infty<\eta<0$
and $0\le r<\infty$,
 this being mapped to 
$-\pi/2<\eta'<0$, $0\le r'< \pi/2$ and $\eta'-r'>-\pi/2$. 
The boundary
$\eta=0$ is mapped to $\eta'=0$, corresponding to future null infinity. The boundary 
$\eta=-\infty$ with $r<\infty$ is mapped to
$(r',\eta')=(0,-\pi/2)$, corresponding to a big-bang singularity.
The boundary $r=\infty$ with $\eta<\infty$ is mapped to $(r',\eta')=(\pi/2,0)$, 
corresponding to spatial infinity.
The boundary $-\eta=r=\infty$ is mapped to $\eta'-r'=-\pi/2$, corresponding 
to a big-bang singularity.
The trapping horizon terminates at $\eta'=r'=0$.
There is no particle horizon but a cosmological event horizon.

\item NP4: $w<-1$\\
In this case, as $a$ increases, the Friedmann equation is dominated by 
the density term and the solution ends with
a future big-rip singularity at a finite value of $t$.
This can be seen from Eqs.~(\ref{eq:tilde_variables}) and (\ref{eq:tilde_variables_NP}).
As $a$ decreases, the density term becomes subdominant 
compared to the  curvature term. 
The scale factor $a$ vanishes at $t=0$, with
$\dot{a}$ approaching unity and
the curvature invariants vanishing in this limit.
This implies that the solution approaches the Milne universe as 
$t\to 0$ and is extendible beyond the $t=0$ null hypersurface. 
The domain of $\eta$ and $r$ is 
$-\infty<\eta<0$ and $0\le
      r<\infty$.
This is mapped to the intersection of  
$-\pi/2<\eta'<0$, $0\le r'< \pi/2$ and $\eta'-r'>-\pi/2$.
The conformal diagram is similar to the NP2 case, except that
$\eta'=0$ corresponds to a future big-rip singularity, while  
$\eta'-r'=-\pi/2$ corresponds to a regular null hypersurface, 
which is $C^{2}$-extendible to Minkowski spacetime.
However, note that 
$\rho_{0}$ in Eq.~(\ref{eq:integration_conservation}) must 
be constant in the whole spacetime
to ensure the conservation law (\ref{eq:energy_equation}).
Since $\rho_{0}$ trivially vanishes in Minkowski spacetime,
the $C^{2}$-extension to Minkowski breaks the 
conservation law across the null hypersurface.
The trapping horizon terminates at $\eta'=r'=0$ and is spacelike.
There is a cosmological event horizon but no particle horizon. There are then two subcases.

\begin{itemize}
 \item NP4a: $-5/3\le w<-1$\\
In this case, $\eta=0$ corresponds to future null infinity, while
       $\eta=-\infty$ is at a finite affine length.
The boundary $\eta=0$ is both a future big-rip singularity and 
future null infinity.
 \item NP4b: $w<-5/3$\\
In this case, the boundaries $\eta=0$ and 
       $\eta=-\infty$ are both at finite affine lengths.
The boundary $\eta=0$ is a future big-rip singularity
but not future null infinity.
\end{itemize}
\end{itemize}

\begin{table}[htbp]
\begin{center}
\caption{\label{table:negative_curvature} Negative-curvature FLRW
 solutions with $\rho>0$}
\begin{tabular}{llccccccccccc}
\hline\hline
&&~~~$t$~~~&~~~$\tilde t$~~~&~~~$\eta$~~~&~~~$\tilde \eta$~~~&~~~$\eta'$~~~&~~~$a$~~~&~~~$ \tilde a$~~~&~~~$\rho$~~~&~~~$r<\infty$~~~&$r=\infty$
\\
\hline
NP1: &$ -1/3<w<\infty$&$\infty$&$\infty$&$\infty$&$\infty$&$\pi/2$&$\infty$&$\infty$&$0$&$i^+$&$\mathscr{I}^+$
\\
&$$&$0$&$0$&$0$&$0$&$0$&$0$&$0$&$\infty$&BB~\footnote{big bang}&$i^0$
\\
\hline
NP2: &$w=-1/3$&$\infty$&--&$\infty$&--&--&$\infty$&--&$0$&$i^+$&$\mathscr{I}^+$
\\
&$$&$0$&--&$-\infty$&--&--&$0$&--&$\infty$&BB&BB
\\
\hline
NP3: &$-1< w <-1/3$&$\infty$&$0$&$0$&$0$&$0$&$\infty$&$0$&$0$&$\mathscr{I}^+$&$i^0$
\\
&&$0$&$\infty$&$-\infty$&$\infty$&$-\pi/2$&$0$&$\infty$&$\infty$&BB&BB
\\
\hline
NP4a&$-5/3\leq w <-1$&$t_0$&$0$&$0$&$0$&$0$&$\infty$&$0$&$\infty$&FBR~\footnote{future big rip} \& $\mathscr{I}^+$&$i^0$
\\
&&$0$&$\infty$&$-\infty$&$\infty$&$-\pi/2$&$0$&$\infty$&$0$&regular &RNHS~\footnote{regular null hypersurface}
\\
\hline
 NP4b: &$-\infty<w<-5/3$&$ t_0$&$0$&$0$&$ 0$&$ 0$&$ \infty$&$ 0$&$ \infty$&FBR&$i^0$
\\
&&$0$&$ \infty$&$ -\infty$&$ \infty$&$ -\pi/2$&$ 0$&$ \infty$&$0$&regular &RNHS
\\
\hline\hline
\end{tabular}
\end{center}
\end{table}

\subsection{Negative energy density}
For $\rho<0$ and $w>-1/3$, the Friedmann equation implies that
$a$ begins at $\infty$, decreases to 
$a'_{c}:=(-\tilde{a}_{c})^{1/(1+3w)}$ and then
increases to $\infty$.
For $w<-1/3$, it  begins at $0$, increases to $a'_{c}$ and then
decreases to $0$.
More precisely,
for $w\ne -1/3$, 
integrating Eq.~(\ref{eq:Friedmann_dust}) gives
\begin{eqnarray}
 \tilde{a}&=&\tilde{a}'_{c}
  \frac{1+\cosh\tilde{\eta}}{2},~~\tilde{t}=\tilde{a}'_{c}
  \frac{\tilde{\eta}+\sinh\tilde{\eta}}{2} \, , 
\label{eq:tilde_variables_NN}
\end{eqnarray}
where $\tilde{a}_{c}'=-\tilde{a}_{c}$, 
$-\infty<\tilde{\eta}<\infty$ and
the conformal time is 
\begin{equation}
 \eta=\frac{1}{1+3w}\tilde{\eta} \, . 
\end{equation}
For $\tilde{\eta}\to \pm \infty$, 
we have $\tilde{a}\to \infty$, with
$\tilde{a}$ bouncing from contraction to expansion
at $\tilde{a}'_{c}$ and $\tilde{\eta}=0$.
The form of $a(t)$ is shown in Fig.~\ref{scalefactor}.
The Misner-Sharp mass $m$ is negative if the energy density is negative. 
Since one always has $2m/R<1$ in this case, as seen from
\begin{equation}
 \frac{2m}{R}=-\frac{\sinh^{2}r}{\cosh^{2}(\tilde{\eta}/{2})} \, ,
\end{equation}
there are no trapped
or marginally trapped spheres.

\begin{figure}[htbp]
\begin{center}
\subfigure[\label{fg:ncndwgm13} NN1: $w> -1/3$]{\includegraphics[height=0.28\textheight]{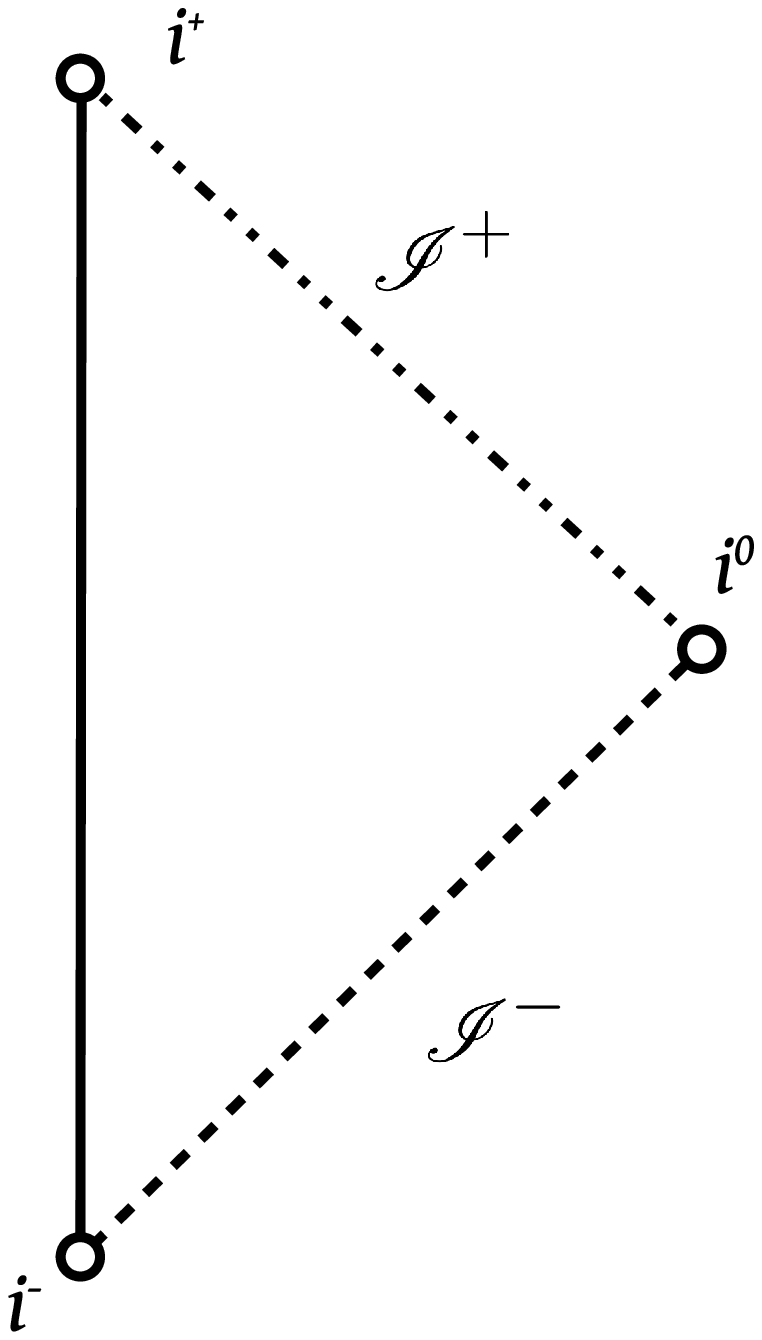}}
\hspace{20mm}  
\subfigure[\label{fg:ncndwm13} NN2: $w=-1/3$]{\includegraphics[height=0.28\textheight]{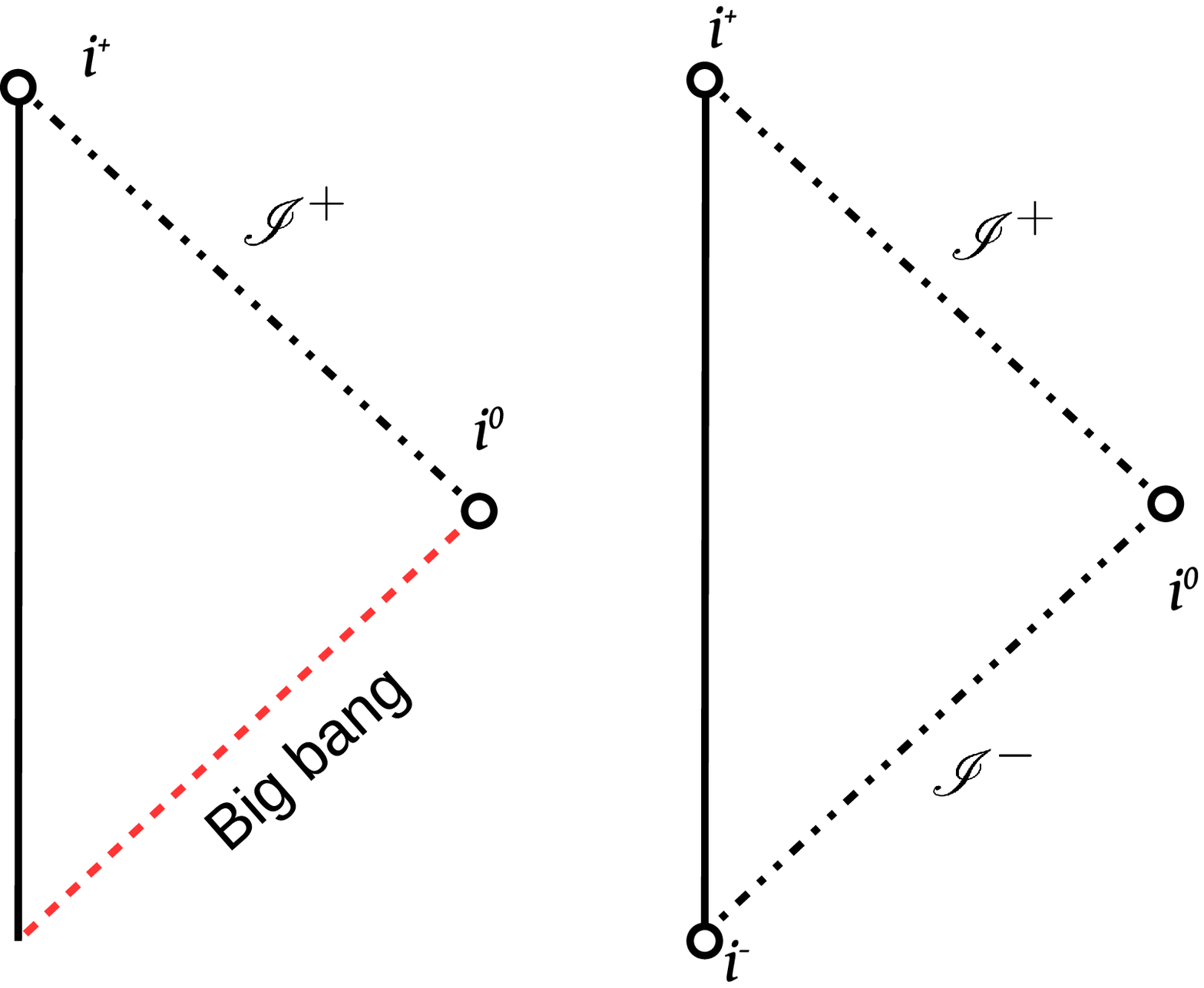}}
\end{center}
\begin{center}
\subfigure[\label{fg:ncndwgm1lm13}NN3:~$-1<w< -1/3$]{\includegraphics[height=0.28\textheight]{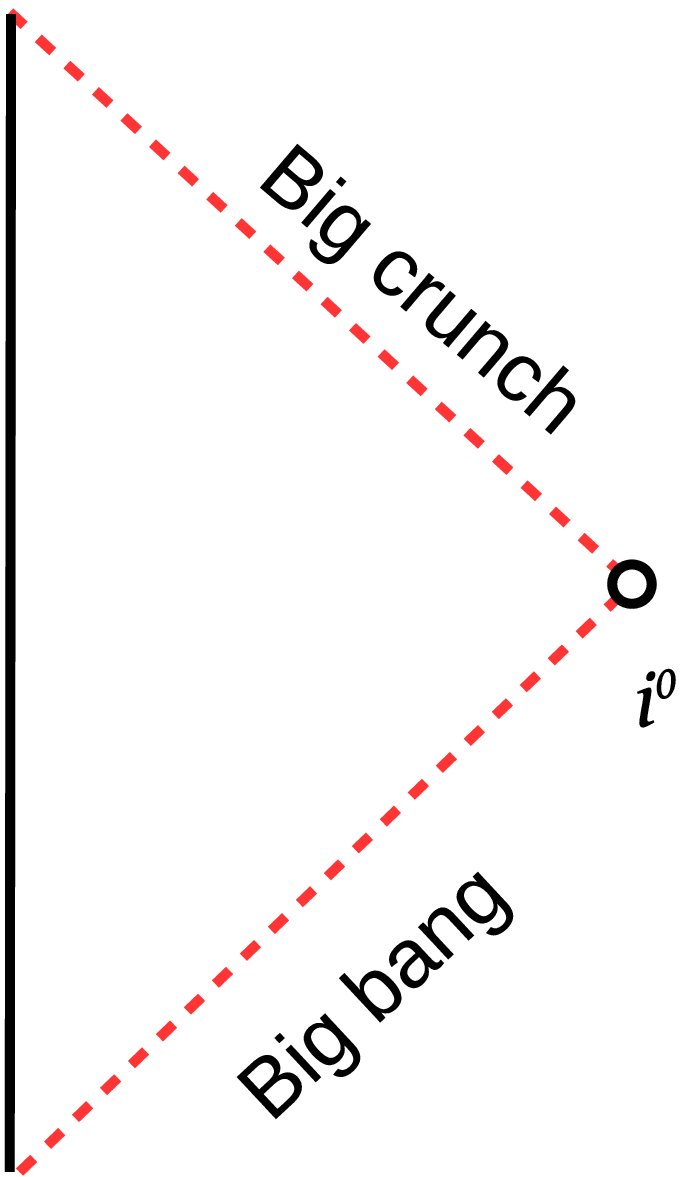}}
\hspace{20mm}  
\subfigure[\label{fg:ncndwlm1} NN4: $w<-1$]{\includegraphics[height=0.28\textheight]{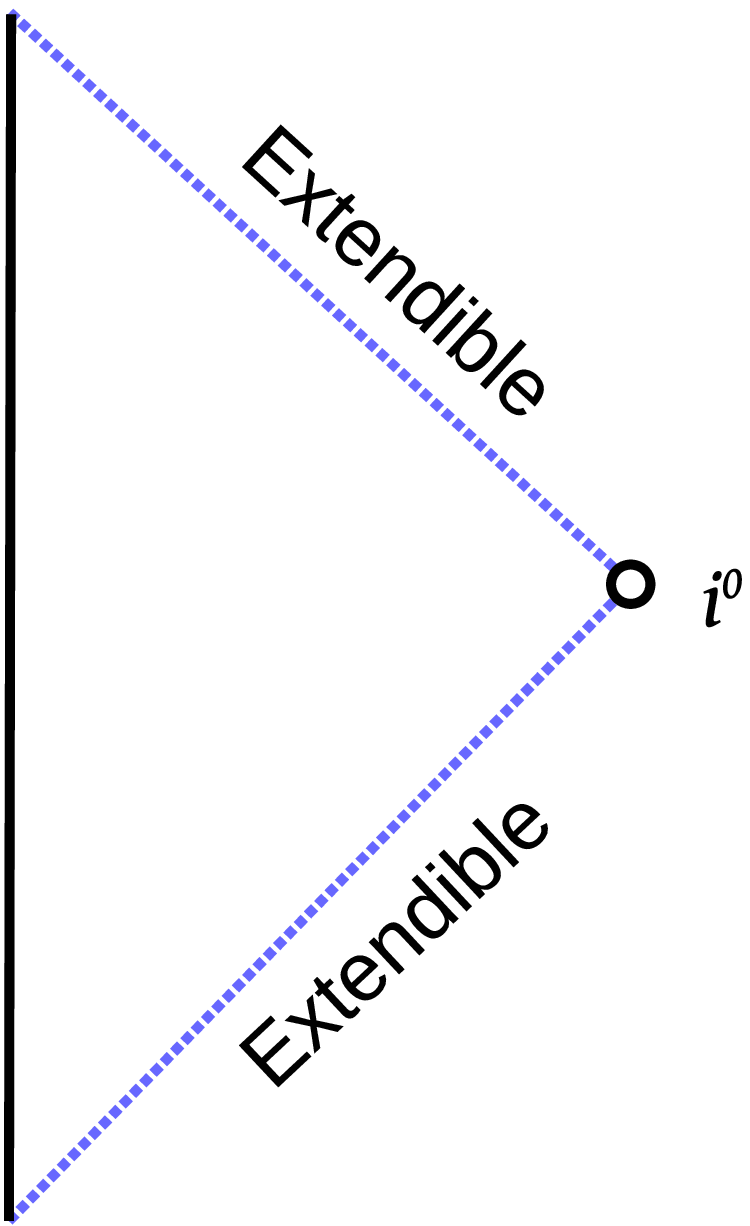}} 
\caption{\label{fg:negative_curvature_negative_density}
The conformal diagrams for the negative-curvature FLRW
 solutions with a negative energy density. 
There are no trapped regions in this case.
Figure~\ref{fg:ncndwgm13} 
corresponds to NN1 ($w>-1/3$).
Figure~\ref{fg:ncndwm13} corresponds to NN2 ($w=-1/3$) with 
the left and right panels applying for 
$0<\tilde{a}'_{c}<1$ and $\tilde{a}'_{c}=1$, respectively.
Figures~\ref{fg:ncndwgm1lm13} and \ref{fg:ncndwlm1} 
correspond to
NN3 ($-1<w<-1/3$) and NN4 ($w<-1$), respectively.
The blue short-dashed lines for $w<-1$ denote 
regular null hypersurfaces
at finite affine lengths, 
beyond which the spacetime is 
at least $C^{2}$-extendible.}
\end{center}
\end{figure}

For $w\ne -1/3$, the affine 
parameter along null geodesics 
for $\eta \to \pm\infty $ is also given 
by Eq.~(\ref{eq:nc_null_affine})
up to an affine transformation. 
The spacetime boundaries are summarised in
Table.~\ref{table:negative_curvature_negative_density} and 
the conformal diagrams shown in
Fig.~\ref{fg:negative_curvature_negative_density} for these cases.
We now discuss the various possibilities.
Note that there are no trapped regions in this case and 
that there is no particle horizon or cosmological event horizon.

\begin{itemize}
 \item NN1: $w>-1/3$\\
In this case, the boundaries $\eta= \pm \infty$ correspond to $a= \infty$.
Equations~(\ref{eq:tilde_variables}) and (\ref{eq:tilde_variables_NN}) imply that
these boundaries correspond to infinities at $t= \pm \infty$.
The boundaries $\eta=r=\infty$ and 
$-\eta=r=\infty$ are future and past null infinities,
respectively, while $\eta=\infty$ and $\eta=-\infty$ with $r<\infty$
are future and past 
timelike infinities, respectively. $r=\infty$ with $|\eta|<\infty$
corresponds to spatial infinity. There is no singularity and the 
spacetime is geodesically complete. 

\item NN2: $w=-1/3$\\
In this case, from Eq.~(\ref{eq:Friedmann3}), we 
have $0\le \tilde{a}'_{c}\le 1$, with 
$\tilde{a}'_{c}=0$ reducing to the vacuum case.
For $0 < \tilde{a}'_{c}<1$, 
the Friedmann equation gives the expanding solution
$
 a=\sqrt{1-\tilde{a}'_{c}} \, t,
$
for $0<t<\infty$.
The collapsing branch is just the time reverse of this.
The conformal time is 
$
 \eta= ({1}/{\sqrt{1-\tilde{a}'_{c}}})\ln t,
$
so $-\infty<\eta<\infty$ and
$t=0$ or $\eta=-\infty$ corresponds to a big-bang singularity.
For $\tilde{a}_{c}'=1$, 
$a=a_{0}$ (constant) 
and the resulting metric is 
a static spatially negative-curvature universe. The conformal time is 
$\eta=t/a_{0}$, so $-\infty<\eta<\infty$ and this is conformal to the
Einstein static model. There is no singularity and
\begin{equation}
 \frac{2m}{R}=-\tilde{a}'_{c}\sinh^{2}r \, .
\end{equation}
This is always negative, so there are no trapped regions.
The affine parameter along null geodesics is 
\begin{equation}
 \lambda= \left\{\begin{array}{cc}
  e^{2\sqrt{1-\tilde{a}_{c}'}\eta} & (0<\tilde{a}_{c}'<1)\\
  \eta & (\tilde{a}_{c}'=1)
	 \end{array}\right.
\end{equation}
up to an affine transformation. Therefore, 
for $0<\tilde{a}_{c}'<1$, 
$\eta=\infty$ and $\eta=-\infty$ correspond to null infinity and
the boundary at a 
finite affine length, respectively, while
for $\tilde{a}_{c}'=1$,  
the boundaries $\eta=\pm r= \pm\infty$ correspond to null infinities.

\item NN3: $-1<w<-1/3$\\
In this case, the boundaries $\eta= \pm \infty$ correspond to $a= 0$.
Equations~(\ref{eq:tilde_variables}) and (\ref{eq:tilde_variables_NN})
      imply that $\eta=\infty$ and $\eta=-\infty$ 
correspond to $t=0$ and $t=t_{0}$, respectively, where $0<t_{0}<\infty$.
The energy density $\rho$ diverges at $t=0$ and $t=t_{0}$,  
implying big-bang and big-crunch singularities,
respectively. These singularities are null. 

\item NN4: $w<-1$\\
In this case, the boundaries $\eta= \pm \infty$ correspond to $a= 0$. 
Equations~(\ref{eq:tilde_variables}) and (\ref{eq:tilde_variables_NN}) imply that
they correspond to $t=0$ and $t=t_{0}$, respectively, where
      $0<t_{0}<\infty$. The energy density vanishes both at $t=0$ and $t=t_{0}$.
The Friedmann equation 
is dominated by the curvature term, so it asymptotically approaches the
      Milne one. 
There is no divergence
in the curvature invariants, which 
vanish 
at both $t=0$ and $t=t_{0}$.
Thus $\eta=\pm \infty$ correspond to 
regular null hypersurfaces 
at finite affine lengths. 
The spacetime is 
$C^{2}$-extendible to Minkowski spacetime
beyond the regular null hypersurfaces. 

\end{itemize}

\begin{table}[htbp]
\begin{center}
\caption{\label{table:negative_curvature_negative_density} 
Negative-curvature FLRW solutions with $\rho<0$}
 \begin{tabular}{llccccccccccc}
\hline\hline
&&~~~$t$~~~&~~~$\tilde t$~~~&~~~$\eta$~~~&~~~$\tilde \eta$~~~&~~~$\eta'$~~~&~~~$a$~~~&~~~$ \tilde a$~~~&~~~$\rho$~~~&~~~$r<\infty$~~~&$r=\infty$
\\
\hline
NN1: &$ -1/3<w<\infty$&$\infty$&$\infty$&$\infty$&$\infty$&$\pi/2$&$\infty$&$\infty$&$0$&$i^+$&$\mathscr{I}^+$
\\
&&$-\infty$&$-\infty$&$-\infty$&$-\infty$&$-\pi/2$&$\infty$&$\infty$&$0$&$i^-$&$\mathscr{I}^-$
\\
\hline
NN2: &$w=-1/3$ ($0<\tilde a_c'<1$)&$\infty$&--&$\infty$&--&$ \pi/2$&$ \infty$&--&$0$&$i^+$&$ \mathscr{I}^+$\\
&&$0$&--&$-\infty$&--&$ -\pi/2$&$ 0$&--&$ \infty$&BB~\footnote{big bang}&BB
\\
\hline
NN2: &$w=-1/3$ ($\tilde a_c'=1$)&$\infty$&--&$\infty$&--&$ \pi/2$&$a_0$&--&$ \rho_0$&$i^+$&$ \mathscr{I}^+$
\\
&&$-\infty$&--&$-\infty$&--&$ -\pi/2$&$ a_0$&--&$ \rho_0$&$ i^-$&$ \mathscr{I}^-$
\\
\hline
NN3: &$-1<w<-1/3$&$t_0$&$ -\infty$&$ \infty$&$-\infty$&$ \pi/2$&$0$&$ \infty$&$\infty$&BC~\footnote{big crunch}&BC
\\
&&$ 0$&$ \infty$&$ -\infty$&$ \infty$&$ -\pi/2$&$0$&$\infty$&$
				      \infty$&BB &BB
\\
\hline
NN4: &$-\infty<w<-1$&$ t_0$&$-\infty$&$ \infty$&$ -\infty$&$
			  \pi/2$&$0$&$\infty$&$ 0$&regular &RNHS~\footnote{regular null hypersurface}
\\
&&$ 0$&$\infty$&$ -\infty$&$ \infty$&$-\pi/2$&$ 0$&$\infty$&$0$&regular &RNHS
\\
\hline\hline
\end{tabular}

\end{center}
 
\end{table}

\section{Conclusions}
\label{sec:conclusions}
We have completely classified the FLRW solutions of the Einstein equation with 
the equation of state $p=w\rho$ according to their conformal structure, going beyond
the usual energy
conditions. 
We have classified
the cases in terms of the spatial curvature $K$,
the equation of state parameter $w$ and the sign of the energy density $\rho$.
For the vacuum case, the FLRW solution is 
Minkowski spacetime, while for $w=-1$ it is  the 
de Sitter spacetime for $\rho>0$ and the anti-de Sitter spacetime for $\rho<0$.
For other cases, there is a rich variety of structures. 

The important features are 
as follows: For each spatial curvature, 
the causal nature of the spacetime is the same  for $w>-1/3$, as is 
well known.
For $w=-1/3$, the speed of the cosmological expansion is constant.
For $-1<w<-1/3$, there is a null big-bang singularity for
$K=0$ and $-1$,
but  the solution describes a bouncing universe for $K=1$.
For $w<-1$, the causal structure is rather exotic. 
For $w<-1$ and $K=0$, the universe gradually expands from a vanishing
scale factor
for an infinitely long time and ends with a future big-rip singularity. 
For $w<-1$ and $K=1$, the universe begins with a past big-rip singularity,
contracts, bounces and ends with a 
future big-rip singularity. 
For $w<-1$, $K=-1$ and $\rho>0$, 
the universe emerges from a regular null hypersurface
and ends with a future big-rip singularity. 
For $-5/3\le w< -1$, null geodesics 
cannot reach the future big-rip singularity within a finite affine length, 
while for $w<-5/3$, they can.
A negative energy density 
is only possible  for $K=-1$, which describes 
a bouncing universe with future and past null infinities for $w>-1/3$, 
a universe beginning with a big-bang singularity and 
ending with a big-crunch singularity
for $-1<w<-1/3$, and a universe emerging from
a regular null hypersuface
and then submerging into another regular hypersurface for $w<-1$.
In general, 
the big-bang and past big-rip singularities are
 followed by past and future trapping horizons, respectively, while 
the big-crunch and future big-rip singularities are preceded by future
and past trapping horizons, respectively. 

Although we have focused on the linear equation of state,
 the generalisation to 
other types of  matter fields is 
interesting. In particular, 
it is important to include both a perfect fluid and a positive or
negative cosmological constant,
not only in the context of 
classification of singularities in FLRW spacetimes 
but also from the cosmological point of view.
Since the dominant term on the right-hand side of
the Friedmann equation should determine the properties of the spacetime
boundaries, the structure of big-bang singularities for $w>-1/3$ and 
that of big-rip singularities for $w<-1$ are unchanged even in the
presence of a cosmological constant, although 
the intermediate dynamics of the scale factor can be greatly changed.
In another paper \cite{CHI}, we expand our analysis to derive the
conformal diagrams for solutions which represent  black holes,
wormholes 
and baby universes 
in a cosmological background.

\acknowledgments

We are grateful to D.~Ida, A.~Ishibashi, 
M.~Kimura, T.~Kokubu, F.~C.~Mena and K.-I. Nakao
for helpful comments and fruitful discussions.
This work was partially supported by JSPS KAKENHI Grant No. JP26400282
(T.H.)
and MEXT-Supported Program for the Strategic Research 
Foundation at Private Universities 2014-2017 (S1411024) (T.I.).
T.H. thanks CENTRA, IST, Lisbon and B.C. thanks 
RESCEU, University of Tokyo for hospitality received durng this work.

\end{document}